\documentclass[structabstract]{aa}  
\usepackage{graphicx}
\usepackage{natbib}
\usepackage{txfonts}
\begin{document}
   \title{Spectral energy distributions of an AKARI-SDSS-GALEX sample of galaxies}


   \author{V.\ Buat\inst{1}\and
   E.\ Giovannoli\inst{1}\and
          T.\ T.\ Takeuchi\inst{2,3}\and
         S.\ Heinis\inst{1}\and
         F.-T.~Yuan\inst{3}\and
         D.\ Burgarella\inst{1}\and 
         S.\ Noll\inst{4}\and
         J. Iglesias-P\'aramo\inst{5,6} }

   \offprints{V. Buat}
   \institute{Laboratoire d'Astrophysique de Marseille, OAMP, Universit\'e Aix-marseille, CNRS, 38 rue Fr\'ed\'eric Joliot-Curie, 13388 Marseille cedex 13, France\\
              \email{veronique.buat@oamp.fr,elodie.giovannoli@oamp.fr, sebastien.heinis@oamp.fr, denis.burgarella@oamp.fr}
         \and
             Institute for Advanced Research, Nagoya University, Furo-cho, Chikusa-ku, Nagoya 464--8601, Japan\\
             \email{takeuchi@iar.nagoya-u.ac.jp}
             \and
Division of Particle and Astrophysical Sciences,
               Nagoya University, Furo-cho, Chikusa-ku, Nagoya 464--8602, Japan\\
               \email{yuan.fangting@g.mbox.nagoya-u.ac.jp}
 \and 
             Institut f\"ur Astro- und Teilchenphysik, Universit\"at Innsbruck, Technikerstr.25/8, 6020 Innsbruck, Austria \\
             \email{Stefan.Noll@uibk.ac.at}
             \and
             Instituto de Astrof\'isica de Andaluc\'{\i}a, Glorieta de la Astronom\'ia,
18008 Granada, Spain
\and  Centro Astron\'omico Hispano Alem\'an, C/ Jes\'us Durb\'an Rem\'on 2-2, 04004
Almer\'ia, Spain  \\
             \email{jiglesia@iaa.es}
                       }

   \date{}

 
   \abstract
   {The nearby universe remains   the best   laboratory to understand the physical properties of galaxies  and is a reference  for  any comparison  with high redshift observations. The all sky (or very large) surveys that have been performed from the ultraviolet (UV) to the far-infrared (far-IR) provide  us with large datasets of very large wavelength coverage  to perform  a reference study.}
 {We  investigate the dust attenuation characteristics, as well as the  star formation rate (SFR) calibrations of  a sample of nearby galaxies observed over 13 bands from  0.15 to 160 $\mu$m.  }
{ A sample of 363 galaxies is built from the  AKARI /FIS all sky survey  cross-correlated with  SDSS and GALEX  surveys. Broad band spectral energy distributions are fitted with  the CIGALE code optimized to analyse variations in dust attenuation curves and SFR measurements and based on an energetic budget between the  stellar and dust emission.}
{Our galaxy sample is primarily selected in far-IR and   mostly constituted of massive, actively star-forming galaxies. There is some evidence for  a dust attenuation law that is  slightly steeper than that used for  starburst galaxies but  we are unable to constrain the presence or not of a bump at 220 nm.   We confirm that a time-dependent dust attenuation is necessary to perform the best fits. Various calibrations of the dust attenuation in the UV  as a function of UV-optical colours  are discussed. A calibration of the current SFR  combining UV and total IR emissions is   proposed  with an  accurate estimate of dust heating by old stars. For the whole sample, 17$\%$ of the total dust luminosity  is unrelated to the recent star formation. }
   {}

\titlerunning{AKARI-SDSS-GALEX galaxies  }
   \maketitle

%

\section{Introduction}
The broad-band spectral energy distribution (SED) of a galaxy represents  the combination of the emission from both stars of all ages and interstellar dust that interact in a complex way by means of the  absorption and scattering of the stellar light by dust grains.  The stars emit from the UV to the near-IR whereas the mid and far-IR emission comes from interstellar dust heated by the stellar emission. 
The broad-band SEDs can be reconstructed from a library of stellar tracks by assuming a star formation history, an initial mass function, and a scenario for dust attenuation to reproduce the observed UV-optical distribution and  the re-emitted emission in mid and far-IR. 
By comparing  data with models, one can attempt to derive  some physical parameters related to the   star formation history and dust attenuation in a homogeneous way and simultaneously for all galaxies of a given sample. Practically  the current star formation rate (SFR), the stellar mass, and  the amount of dust attenuation are commonly derived   from broad-band photometry \citep[e.g.][and references therein]{salim07, dacunha08, walcher08, walcher10}.  Without spectral information and given the high degree of degeneracy in the SEDs, details about the star formation history such as   stellar population ages,  or  specific aspects of dust attenuation such as  its wavelength dependence, are  more difficult to constrain \citep[e.g.][]{walcher10, noll09b} .\\
With the availability of mid and far-IR data for large samples of galaxies,  codes that combine stellar and dust emission to analyse SEDs are particularly useful. 
The standard approach consists of solving the radiation transfer in model galaxies to build self-consistent SEDs from the UV to the far-IR  \citep[e.g.][]{popescu00, silva98, tuffs04}. These sophisticated models require complex calculations and are not directly applicable to large samples of galaxies. Libraries of templates can be produced \citep{iglesias07,popescu10}, as well as some recipes \citep{tuffs04}, but the   various  free parameters and physical assumptions are difficult to constrain  with only broad-band, integrated fluxes. Another  tool applicable to large samples of galaxies consists of models based on  a simple  energetic budget where  the global dust emission corresponds to the difference between the emitted and observed stellar light  \citep{dacunha08,noll09b}. Codes based on synthetic stellar population modeling and simple emission properties for the dust are well suited  to analysing large datasets at the expense of an over simplification of the physical processes at work in galaxies. \\
The natural laboratory for this SED analysis is the nearby universe for which plenty of  high-quality photometric data are available.
  
  Several  studies  have been  based on the {\it Spitzer} Nearby Galaxies Survey (SINGS) \citep{kennicutt03}, which provided  broad-band spectral energy distributions  from 0.15 to 160 $\mu$m for  75 galaxies \citep{dale07}. \citet{dacunha08} and \citet{noll09b} illustrate the capabilities of their fitting code with this sample. \\
  In the present work, we investigate the properties of nearby galaxies by selecting a sample of objects that is as large as possible, observed during the  {\it AKARI}, SDSS, and {\it GALEX} surveys that  correspond to 13 photometric bands from 0.15 to 160 $\mu$m.  Particular care was taken to measure total flux densities for the extended sources. Our aim is to  perform SED fitting to deduce physical properties in a homogeneous and consistent way, focusing on dust attenuation and SFR measurements. 
  The originality of our study is to extend the wavelength coverage beyond 100 $\mu$m (the longest wavelength observed by IRAS) so that  we consider only galaxies with a reliable flux at 140 $\mu$m: our selection is close to a selection at 140 $\mu$m and  consists of  363 unresolved objects with a very large wavelength coverage and  secure measurements, allowing a statistical analysis.  A selection at 140 $\mu$m, beyond the peak of dust emission for most galaxies  is particularly valuable for further comparison with distant galaxies observed in sub-mm by {\it Herschel}. It is also the first application of our code CIGALE to a large sample of nearby galaxies with such a wavelength coverage. \\
  In section 3, we  describe the SED fitting method, the reliability of its results derived by  the analysing a mock catalogue and the specific results obtained for our galaxy sample. Section 4 is devoted to the analysis of dust attenuation characteristics and section 5 to SFR calibrations. The conclusion is presented in section 6. All magnitudes are given in AB system.  We assume that $\Omega_m = 0.3$, $\Omega_{\Lambda} = 0.7$, and $ H_0 = 70 {\rm~ km~ s^{-1}~ Mpc^{-1}}$.  


\section{Data}
\begin{table}
\caption{Sample of galaxies, data with an asterix (*) are not used to  fit the SEDs with CIGALE }             
\label{table:1}      
\centering                          
\begin{tabular}{c c c c}        
\hline\hline                 
Survey & Band & Wavelength & Nb of sources \\    
\hline                        
   {\it GALEX}   & FUV & 153 nm &363\\      
   {\it GALEX}   & NUV & 231 nm    & 363\\
   SDSS & u,g,r,i,z & 355, 469, 617, 748, 893 nm &    363 \\
  {\it AKARI}/FIS$^*$ & N60 & 65 $\mu$m & 129 \\
   {\it AKARI}/FIS$^*$& WIDE-S & 90  $\mu$m & 357 \\
   {\it IRAS}& Band-3 & 60 $\mu$m & 359\\
   {\it IRAS}& Band-4 & 100 $\mu$m  & 354\\
   {\it AKARI}/FIS& WIDE-L & 140  $\mu$m &363 \\
   {\it AKARI}/FIS& N160 & 160  $\mu$m& 111\\
    
\hline                                   
\end{tabular}
\end{table}

\begin{figure}
 \includegraphics[width=8cm]{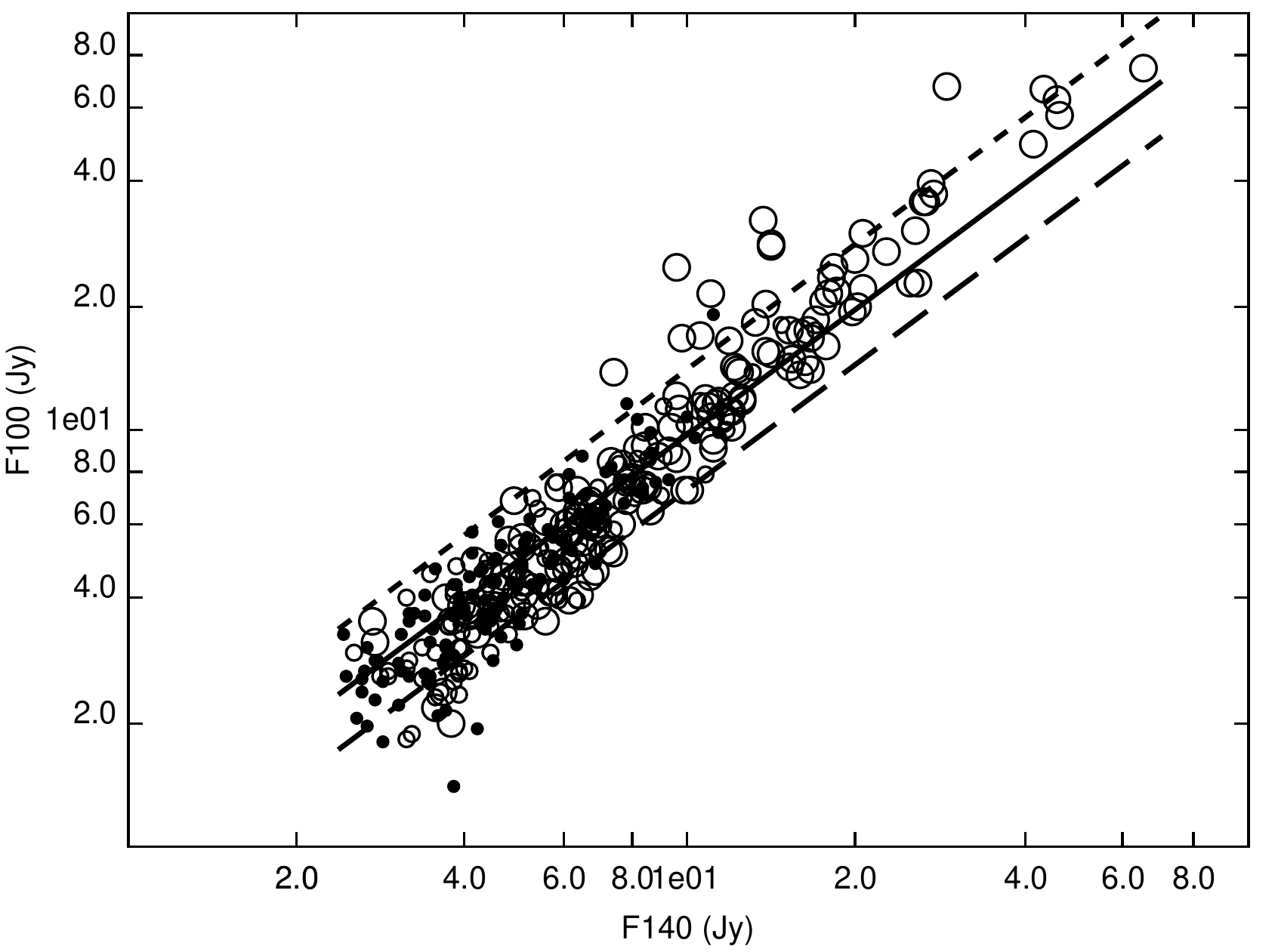}
 \includegraphics[width=8cm]{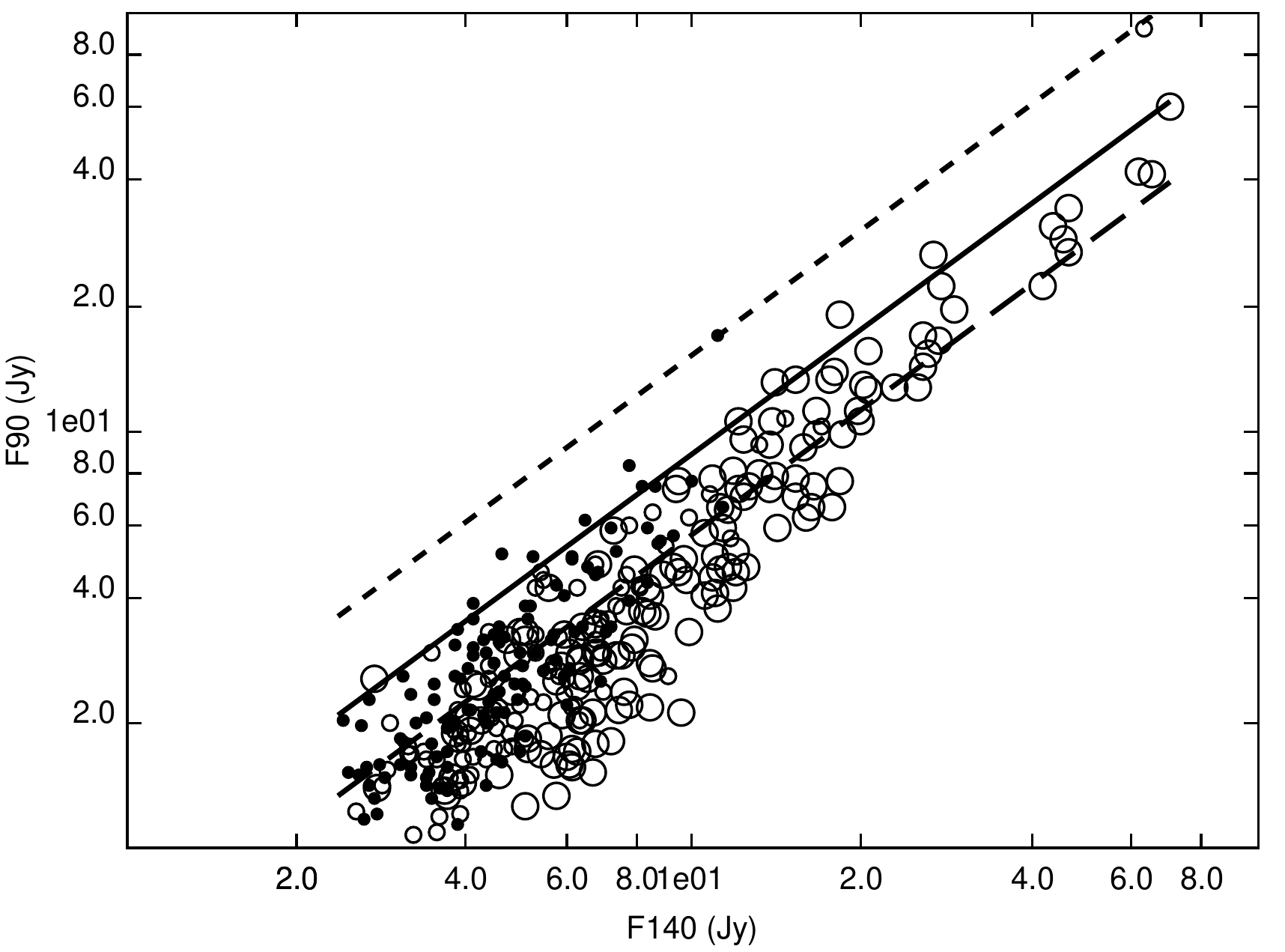}
  \caption{ {\it AKARI}  fluxes at 140 $\mu$m  plotted against {\it IRAS} fluxes at 100 $\mu$m (upper panel) and AKARI fluxes at 90 $\mu$m (lower panel). The sample is divided into  three sub-samples according to the diameter d of the galaxies: d $\rm < 1'$ (dots), $\rm 1'<d<1.5'$ (small circles), and $\rm d > 1.5'$ (large circles). The data are compared to model predictions from \citet{dale02} for three different values of the $\alpha$ parameter, $\alpha=1.5,2$, and 2.5 (dotted,solid, and dashed lines respectively). }
      \label{calib}
\end{figure}

\begin{table*}
 \centering
\begin{tabular}{ l c l}

  \hline
  {Parameters} & {Symbol} & {Range} \\
    \hline
 {Star formation history} & &   \\
  \hline   
metallicities (solar metallicity) & $Z$ &  0.02 \\
 $\tau$ of old stellar population models in Gyr & $\tau_1$ & 1;  3.0; 5.0; 10.0 \\ 
ages of old stellar population models in Gyr & $t_1$ & 13  \\
ages of young stellar population models in Gyr & $t_2$ & 0.025; 0.05; 0.1 ;0.3 ;0.5 ;1.0\\
                                             
fraction of young stellar population & $f_{ySP}$ & 0.001; 0.01; 0.1; 0.999 \\
IMF & K & Kroupa \\

  \hline
  {Dust attenuation} & &  \\
  \hline  
  Slope correction of the Calzetti law & $\delta$ & -0.3; -0.2; -0.1; 0.0; 0.1; 0.2\\
  V-band attenuation for the young stellar population & $A_{V,ySP}$ & 0.15; 0.30; 0.45; 0.60; 0.75; 0.90; 1.05; 1.20; 1.35;  \\
                                       &                &1.5; 1.65; 1.8; 1.95; 2.1 \\
  
  Reduction of $A_{V}$  basic for old SP model & $f_{\rm att}$ & 0.0; 0.50; 1.0 \\
    \hline                                   
 {IR SED} & &\\
    \hline
 IR power-law slope & $\alpha$ & 1.0; 1.5; 1.75; 2.0; 2.25; 2.5; 4.0\\
  \hline   
  \end{tabular}
  \caption{List of the input parameters of the code CIGALE and their selected range.}
  \label{table4}
\end{table*}

 Our aim is to build a galaxy sample with  high  quality fluxes from the UV to the far-IR. 
The present sample is built  from the primary sample of \citet{takeuchi10}.  It consists of galaxies observed  as part of the {\it AKARI}/FIS all sky survey \citep{kawada07} and restricted to the area covered by SDSS/DR7 \citep{abaza} and the  {\it GALEX} All Imaging Survey \citep{martin05} for an additional  cross-match.   \citet{takeuchi10} cross-matched the first primary catalogue of the {\it AKARI}/FIS All Sky Survey  ($\beta$-1 version) with the {\it IRAS}/PSCz \citep{saunders00} to select 776 sources confirmed as galaxies,  with a known redshift  and observed with {\it GALEX}. For 607 of these 776 sources, SDSS images are suitable for performing   accurate  photometry (no bright star superposed). We refer to \citet{takeuchi10} for  details of the sample selection and photometric measurements. Briefly, specific aperture photometry was performed for each source in the {\it GALEX} and SDSS images at the $AKARI$ positions  to avoid shredding. Total flux densities were measured in elliptical apertures on the $GALEX$ images using the method described in \citet{iglesias07}. When sources were  shredded in  the SDSS images during the deblending step of the pipeline (most of the cases), we used the photometry of the parent object measured before deblending within the Petrosian radius. The errors due to the photometric measurements at UV and optical wavelengths were measured  and are of the order of  $\sim 5 \%$.

  We  collect the  fluxes from the {\it AKARI}/FIS All-Sky Survey Bright Source Catalogue Version 1.0 \citep{yamamura},  whose errors are smaller than those of  the $\beta$-1 version used in \citet{takeuchi10}.  We  restrict the sample to galaxies whose 140 $\mu$m flux is of high quality (FQUAL140=3), 363 sources fulfiling this condition. We find that most of selected sources  (357 sources)  also have  high quality fluxes at 90 $\mu$m but the fraction of these sources with reliable fluxes  at  65 and 160 $\mu$m is about $\sim 1/3$.The fluxes uncertainties are estimated to be $\sim 20\%$ as quoted in  the {\it AKARI}/FIS All-Sky Survey Bright Source Catalogue.  All the 363 sources were detected by $GALEX$ and in  all SDSS bands.\\
In Table 1, we  gather  the number of sources with high quality fluxes at each wavelength of interest, the  sample used for our SED analysis consisting   of the 363 sources reliably  detected at 140 $\mu$m and in all the  SDSS bands. Thirteen bands are considered, most  galaxies being detected in 11 of them. All the fluxes are corrected for Galactic extinction using the \citet{schlegel98} dust maps and the \citet{cardelli89} extinction curve. The redshift distribution of the sources is similar to that of the original sample of \citet{takeuchi10} with a mean value $<z> = 0.016$\\
Since the  detection rate at 65  $\mu$m  is  low (129/363 sources), we prefer to consider the {\it IRAS} fluxes at 60 $\mu$m available for 359 sources (we use the 60 $\mu$m co-added fluxes from the PSCz and consider only sources with an  {\it IRAS} quality flag equal to 3). 
 {\it AKARI} fluxes at  65 and 90 $\mu$m  are  systematically lower than {\it IRAS} fluxes  at 60 and  100 $\mu$m, respectively \citep{takeuchi10}. This discrepancy is likely to be due to the higher spatial resolution of {\it AKARI}  data (pixel size 27 arcsec, FWHM of the PSF equal to 39 arcsec \citep{kawada07}): the point source extraction is no longer valid for slightly extended nearby sources (Yuan et al. in preparation). Therefore, at 100 $\mu$m we also consider {\it IRAS} co-added fluxes for the 354 sources safely detected by {\it IRAS} at 100 $\mu$m (IRAS quality flag equal to 3), in section 3 we discuss the difference in the estimate of the total IR luminosity $L_{\rm IR}$ when using  either only  {\it AKARI}  data or a combination of {\it AKARI} and  {\it IRAS} data.
Optical diameters  measured at the isophotal level B= 25 mag arcsec$^{-2}$ are compiled from the HyperLeda database \footnote{http://leda.univ1-lyon1.fr/}.\\
Any limitation of the  spatial resolution and point source extraction are expected to be less important at 140 and 160 $\mu$m than at 65 and 90 $\mu$m, with a pixel size of 44 arcsec and a PSF FWHM of 58 and 61 arcsec, respectively, at 140 and 160 $\mu$m \citep{kawada07}. In Fig~\ref{calib}, the 140  $\mu$m fluxes are compared to the fluxes at 90  and 100  $\mu$m  together with   the predictions of the  \citet{dale02} models  representative of galaxies  in the nearby universe (corresponding to an $\alpha$ parameter of the \citet{dale02} models between 1.5 and 2.5 as described in section 3) . Models and $IRAS$ data at 100 $\mu$m are found to be consistent without any obvious trend with the galaxy diameters whereas {\it AKARI}  fluxes at 90 $\mu$m appear systematically underestimated, the discrepancy increasing with the optical diameter, confirming the analysis results  of \citet{takeuchi10} and Yuan et al. (in preparation).
 
\section{SED fitting}
\subsection{CIGALE code}
An efficient  way to derive physical parameters of  star formation and dust attenuation  homogeneously   is to fit the observed SED with models from a stellar population synthesis code.
We use the code CIGALE (Code Investigating GALaxy Emission) \footnote {http://www.oamp.fr/cigale}, which derives physical information about galaxies by fitting their UV-to-far-IR SED  \citep{noll09b,giovannoli10}. 
A  Bayesian analysis is used to derive galaxy properties similar to that  developed by \citet{kauffmann03a}. CIGALE combines   a UV-optical stellar SED and  a dust, IR-emitting component. First,  models are built,   then each model is quantitatively compared to the observed SEDs  as we account for the uncertainties in the observed fluxes. The probability function of each parameter is calculated and the estimated value of the parameter and its error correspond to the mean and standard deviation of this distribution. Models are generated with a stellar population synthesis code, assuming a particular star formation history  and  dust attenuation scenario. 
The energetic balance between dust-enshrouded stellar emission and re-emission in the IR is carefully conserved by combining the UV-optical and IR SEDs.   We refer the reader to \citet{noll09b} for a detailed description. 
Briefly, the IR SEDs are built from the  \citet{dale02} templates. CIGALE allows for  additional dust emission from a dust-enshrouded  AGN but does not include the   unobscured emission of an active nucleus.  Therefore, we discard five galaxies classified as Seyfert 1 by \citet{veron06}.
We  then performed several tests after  adding   two  AGN templates (corresponding to a PAH-free emission with  respectively $A_{\rm V} = $ 32 and 64 mag of extinction from the \citet{siebenmorgen07} library)  to the SED fitting \citep{noll09b} and found that the AGN  contribution to the total IR emission never exceeds 12$\%$ (without any difference for galaxies classified as active or not) but is very badly constrained in the absence of mid-IR data (a null contribution of the AGN component is never excluded by the Bayesian analysis).

In the following, we describe the parameters that are  crucial to this study. 
The input parameters values used in this work are listed in Table \ref{table4} and presented in this  section. They were chosen after several trials and represent  a compromise between a good estimate of the parameters of interest and  the duration of the SED fitting process.   An over sampling of ill-constrained parameters may also reduce the accuracy   of well-constrained parameters by  increasing the number of models of moderate probability and consequently the dispersion in the parameter estimates. This effect was found to be moderate but real, thus  there is no need to perform a  rigorous  sampling of all the parameters. The range of values  was chosen to be  large for the first tests, then reduced by excluding values never chosen during the $\chi^2$ minimization. The sampling of the parameters was  then  optimized in terms of computer time and parameter estimation.

\subsubsection{Stellar populations and star formation histories}
  \begin{figure}
\centering
 \includegraphics[width=9cm]{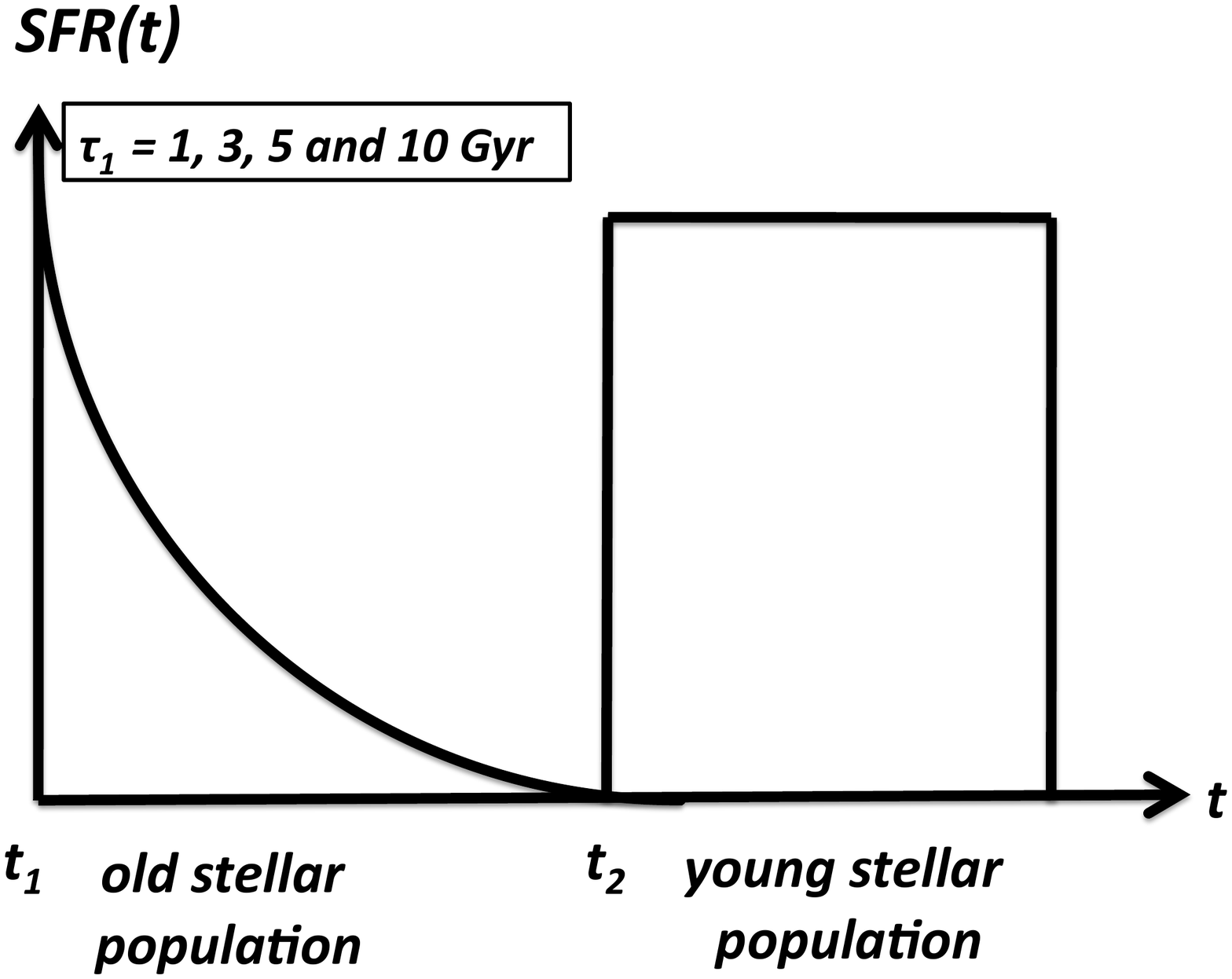}
    \caption{Scenario of star formation history adopted in this work. The old stellar population is produced by an exponentially decreasing star formation rate.  By following a steep-to-moderate e-folding rate $\tau_1$, the young stellar population  is  created  in  $t_2$ years, at a constant, adjustable rate.}
    \label{SFH_2}    
\end{figure}
 \begin{figure}
\centering
\includegraphics[width=9cm]{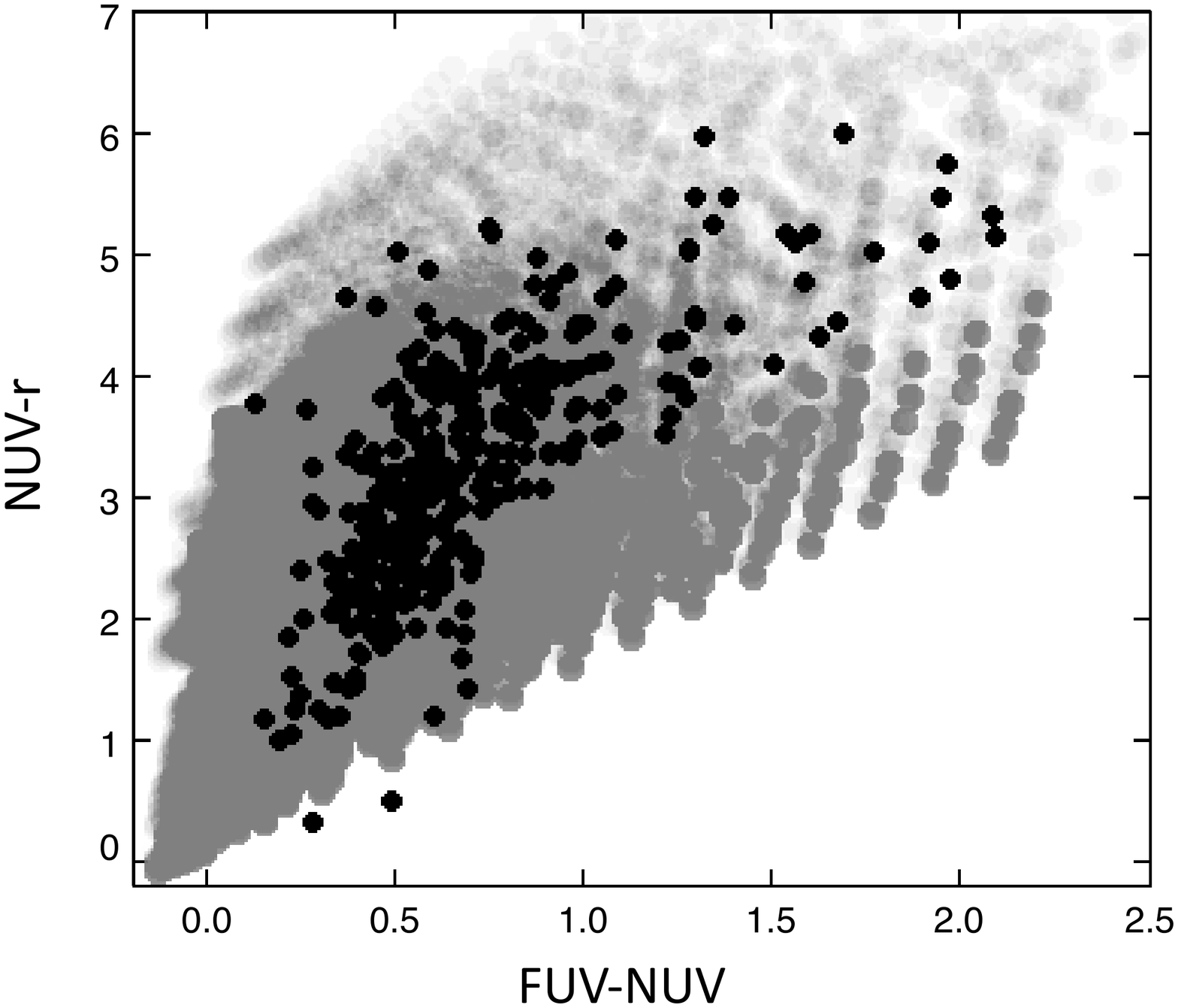}
 \caption{FUV-NUV versus NUV-r colours for observed (black point) and modeled (grey point) data.}
     \label{prior}
\end{figure}
\begin{figure}
\centering
\includegraphics[width=9cm]{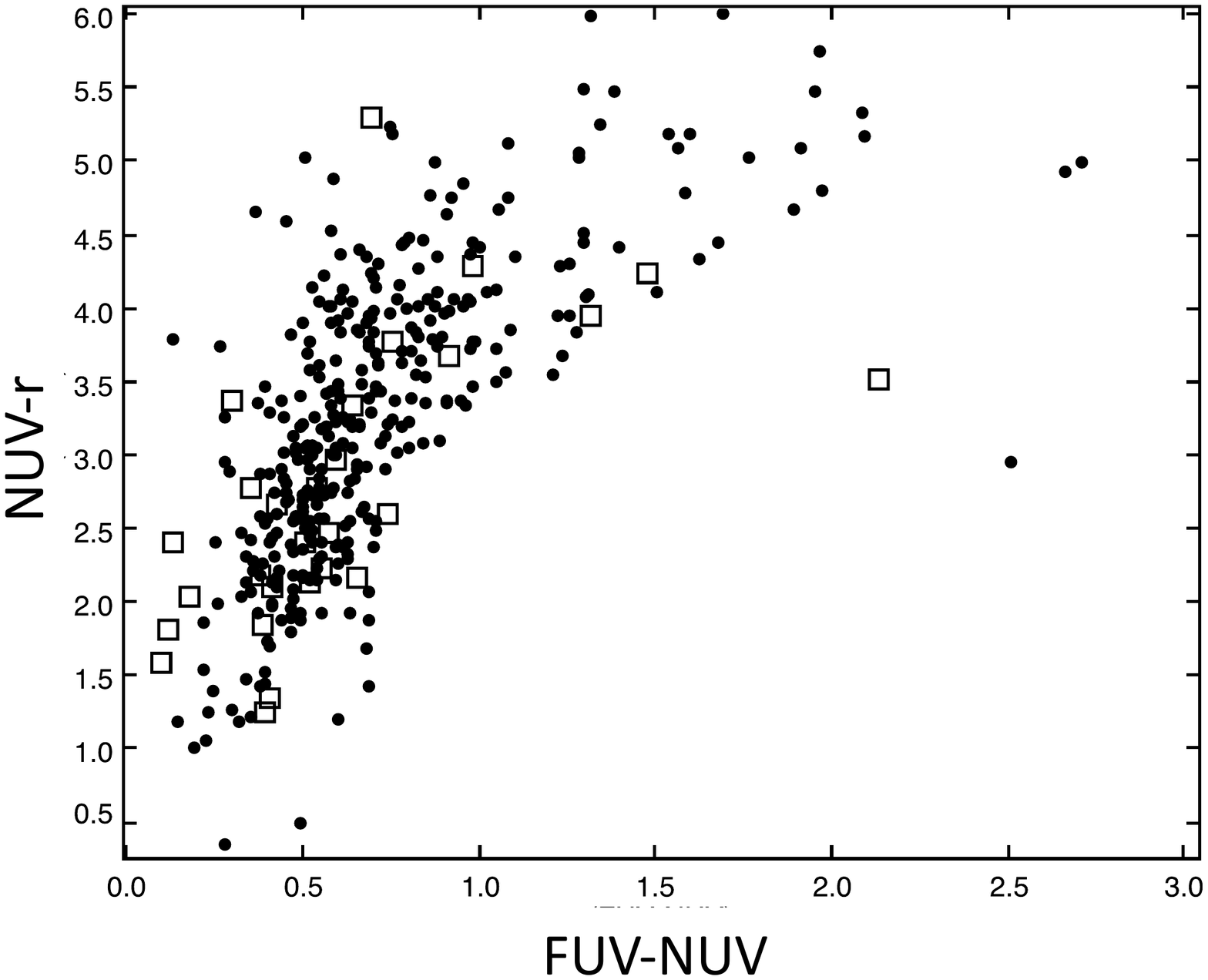}
 \caption{FUV-NUV versus NUV-r colours for observed (black point) and mock  (empty square) data.}
     \label{mock-colours}
\end{figure}
  \begin{figure*}
 \centering
 \includegraphics[width=13cm]{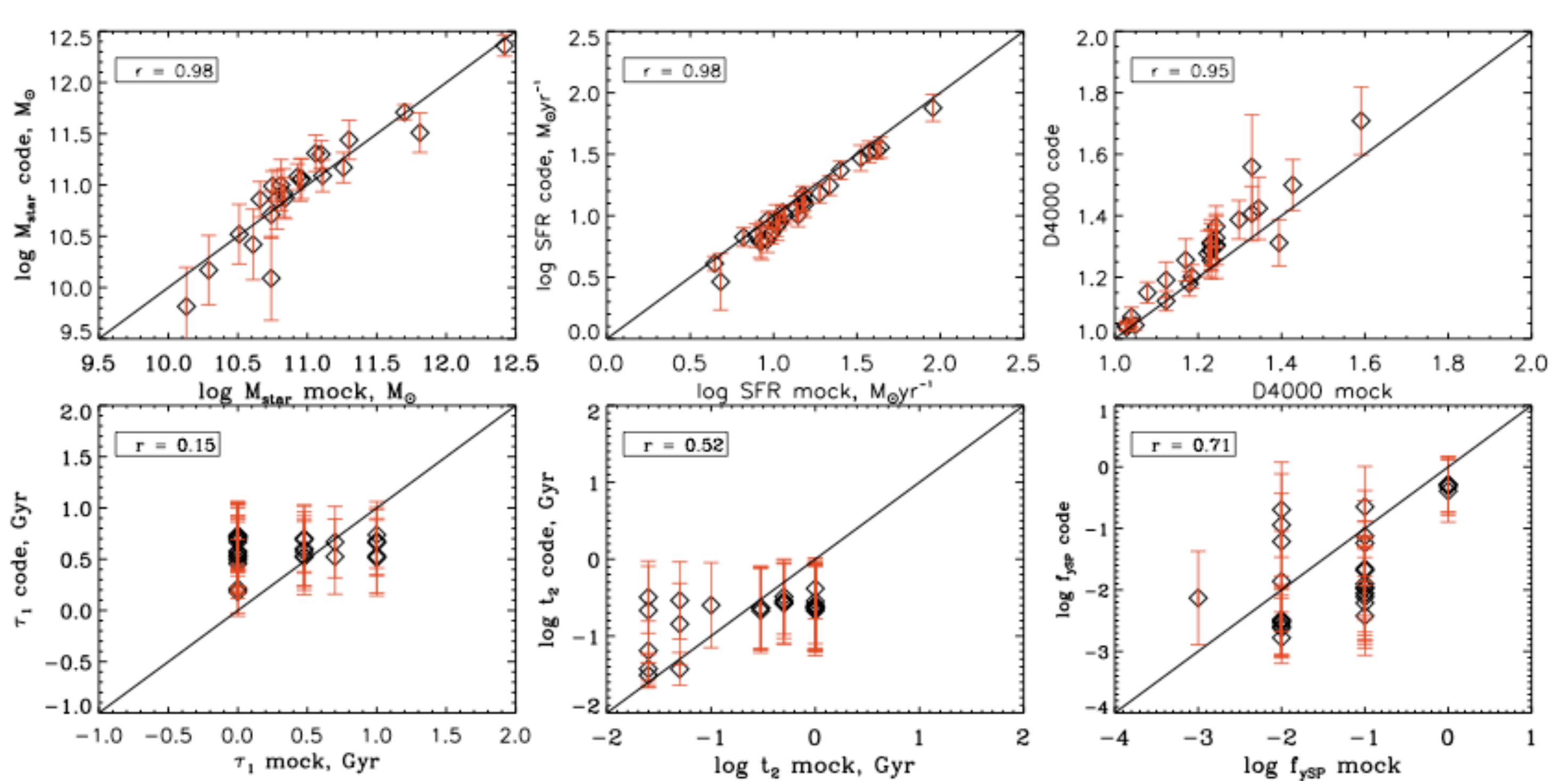}
  \includegraphics[width=13cm]{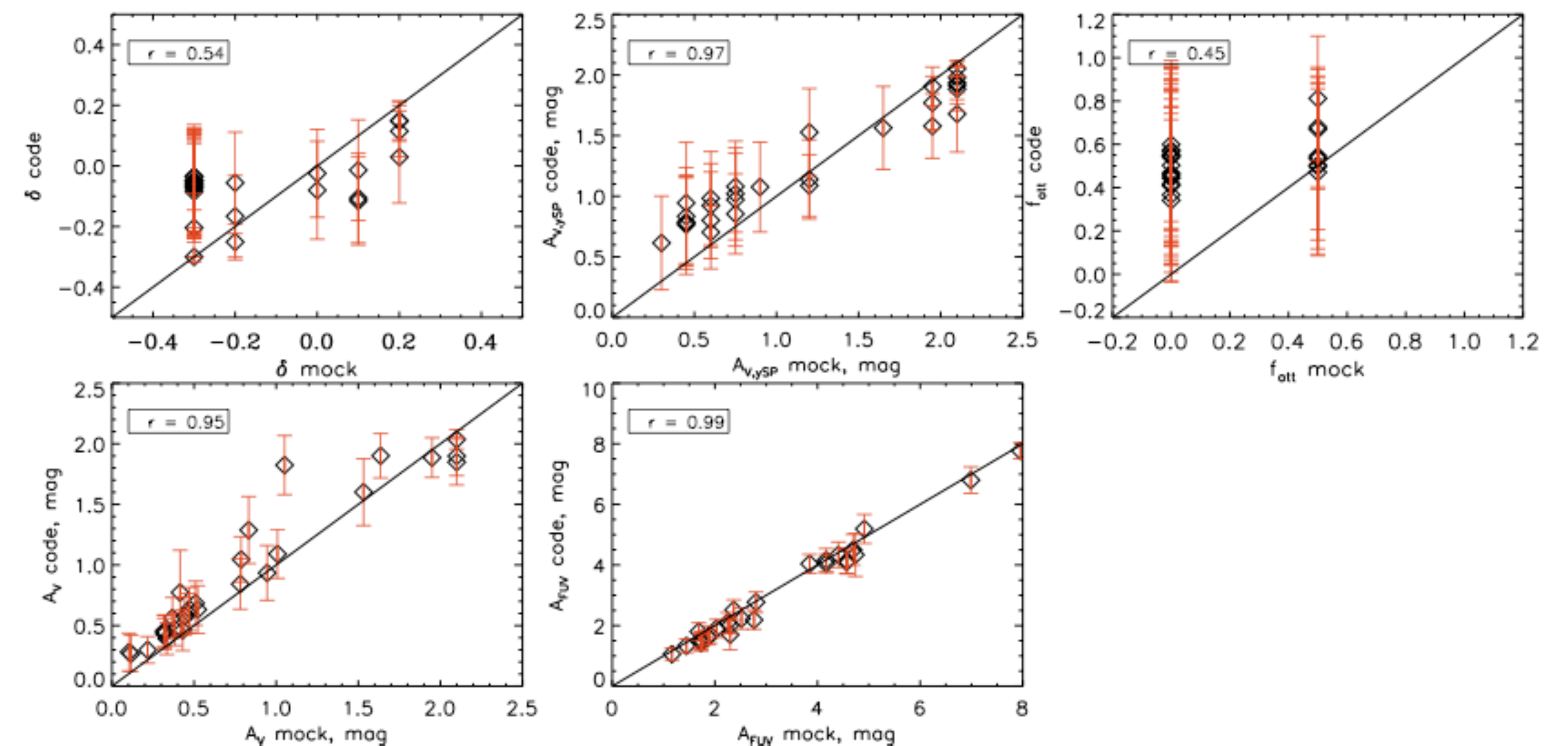}
   \includegraphics[width=9cm]{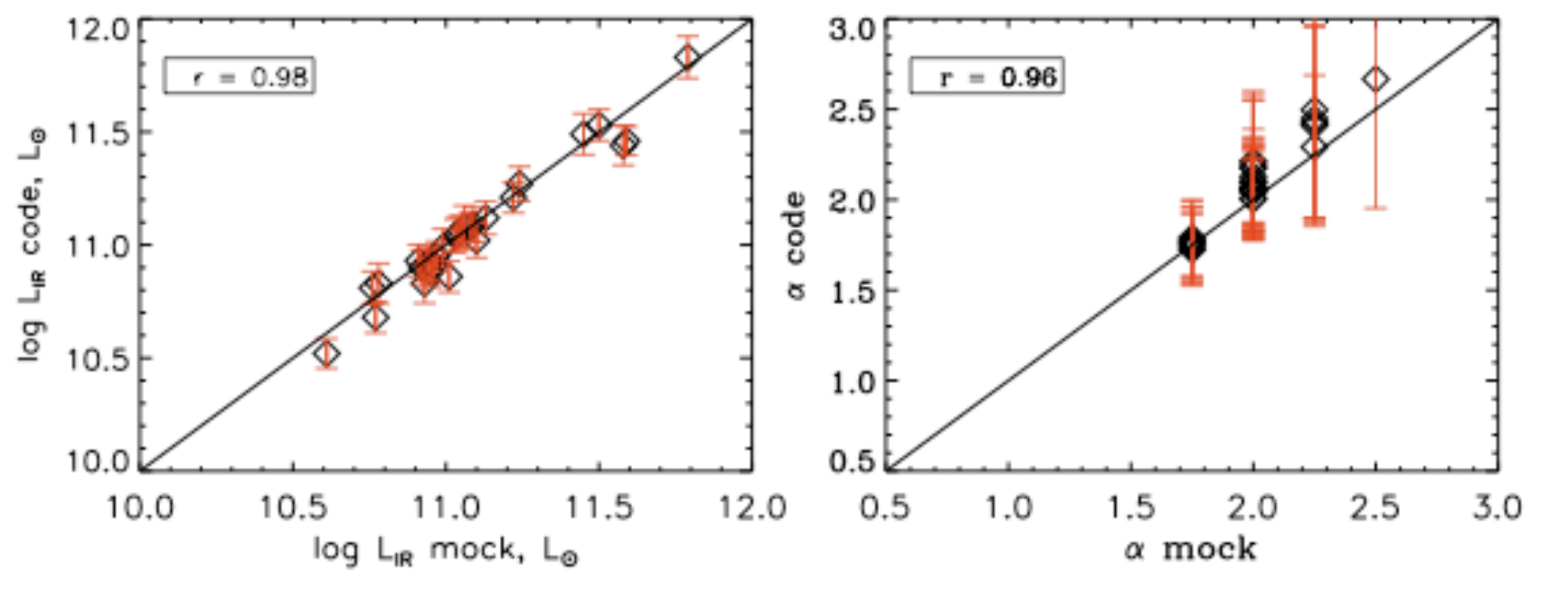}
  \caption{Results of Bayesian analysis of the mock catalog. Each plot corresponds to an output  parameter considered in this work and discussed in the text. The first six  output parameters  are related to the star formation history ($M_{\rm star}$, SFR, D4000, $\tau_1$, $t_2$, and $f_{\rm ySP}$);  the next  five  output parameters characterize the dust attenuation ($\delta$,  $A_{\rm V,ySP}$, $f_{\rm att}$, $A_{\rm V}$, and  $A_{\rm FUV}$) and the two last ones define the IR dust emission ( $L_{\rm IR}$ and $\alpha$). The 'exact' values are plotted on the x-axis, the results applying  the SED fitting method to  the mock data are plotted on the y-axis, and the standard error given by the Bayesian analysis is over plotted as an error bar for each value. The linear Pearson correlation coefficient (r) is indicated on each plot.}
      \label{mock}
     \end{figure*}

We adopt the stellar population synthesis models of    \citet{maraston05},  which include  a full treatment of the thermally pulsating asymptotic giant branch stars. The chosen initial mass function  is that of \citet{kroupa01}. The metallicity is taken to be solar.
 
The star formation history implemented in CIGALE  is the combination of  two stellar components  resembling  an old and a young stellar populations. The two populations roughly represent  a burst of star formation in addition to a more passively evolving stellar component. We verify that including these two populations improves the results of the SED fitting relative  to the case where  only   a single  exponentially decreasing star formation rate is assumed. The reality is certainly   more complex but adding more components may lead to degenerate solutions. The old stellar component is modeled with an exponentially decreasing SFR (with various values of the e-folding rate  $\tau_1$) that  started 13 Gyr ago ($t_1$ = 13 Gyr), the young stellar component consisting of a burst of constant star formation starting later ( $t_2$ Gyr ago)  whose  amplitude is adjustable (Fig. \ref{SFH_2}). 
We adopt four values of    $\tau_1$ (1, 3, 5, or 10 Gyr) and  $t_2$  is taken in the range 0.025 to 1 Gyr. 
 The two stellar  components are linked by their mass fraction, $f_{\rm ySP}$, which  corresponds to the ratio of the mass locked in the young stellar population mass to the total stellar mass. The parameter $f_{\rm ySP}$ varies in the range 0-1,  with a logarithmic scale (linear or logarithmic variations are allowed for the input parameters).
CIGALE estimates the  SFR, defined as (1-$f_{\rm ySP}$)*SFR$_1$ + $f_{\rm ySP}$*SFR$_2$, where SFR$_1$ and SFR$_2$ correspond to the star formation rates of the old and young stellar populations, respectively. In the following, the SFR refers to the formula calculated for the current time. 

In its newest version, CIGALE also   estimates  the  dust-free D4000 break measured in the spectrum of the composite stellar population. The calibration used is that of \citet{balogh99} (see also section 3.4 for a more detailed discussion).

    \begin{figure*}
    \center
 \includegraphics[width=13cm]{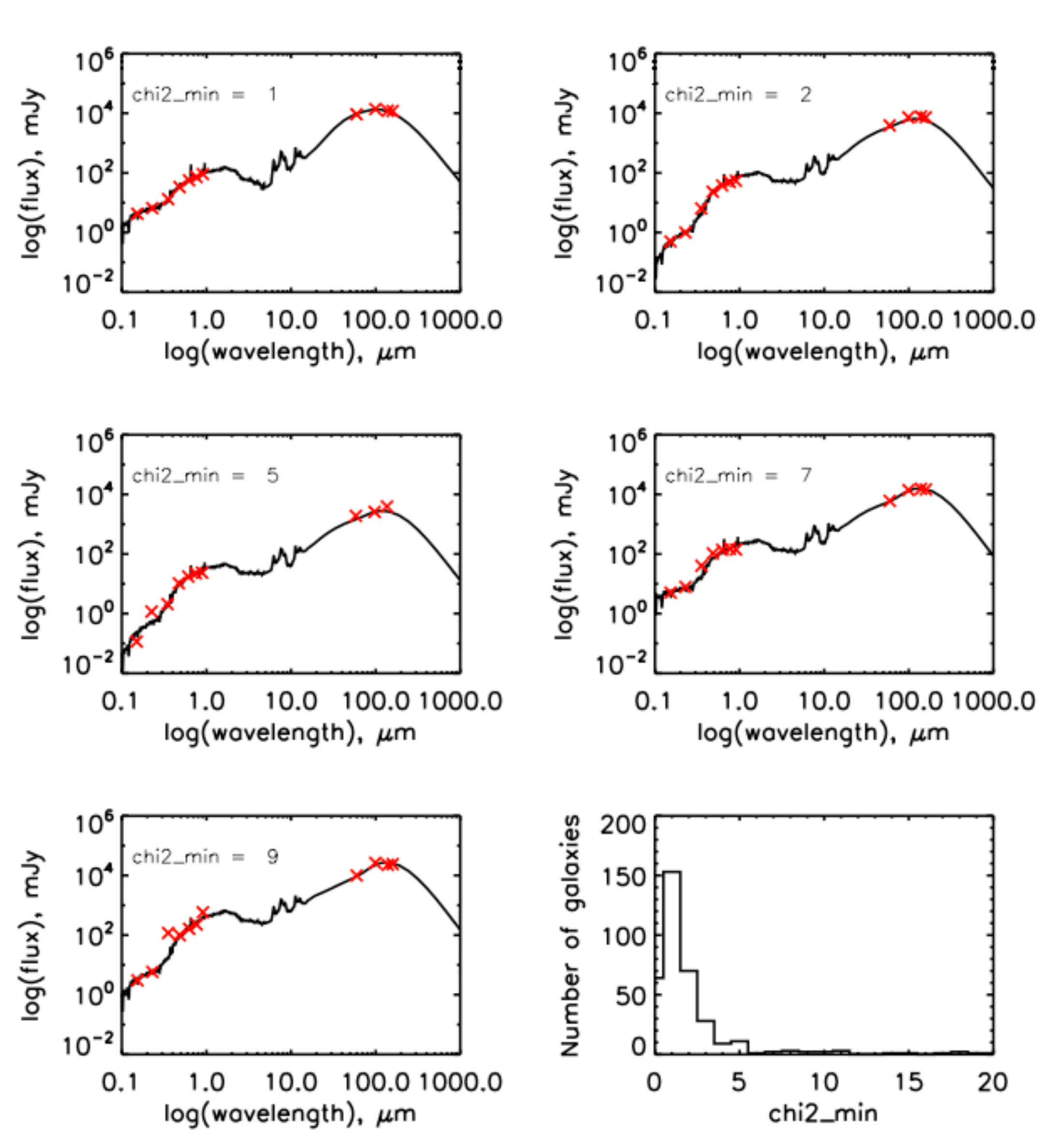}
  \caption{Five examples of best-fit models with different  reduced $\chi^2$ minima  chosen to be representative of our sample. The observed data are plotted with red crosses and the best-fit  model with a solid line. The distribution of the   reduced $\chi^2$ minima for the entire sample  is plotted in the last panel.}
      \label{fits}
     \end{figure*}
 
\subsubsection{Dust attenuation and infrared emission}
To model the attenuation by dust, the code  uses the attenuation law of \citet{calzetti00} as a baseline, and offers the possibility of varying the steepness of the attenuation law and/or adding a bump centred at 220 nm (see \citet{noll09b} for a complete description of the dust attenuation prescription).
To produce different slopes, the parameter $\delta$ is introduced and the original  \citet{calzetti00} law is multiplied by $(\lambda/\lambda_{\rm V})^{\delta}$ with $\lambda_{\rm V} = 550$ nm. We choose $\delta$ between -0.3 and 0.2.  This  range of values was chosen after several trade-offs and we exclude larger departures from the Calzetti law since these solutions  were never chosen by the SED fitting process. Several tests are made to check  whether we are able to constrain the presence of a UV bump with  the available photometric broadbands, including the analysis of mock catalogues,  described in section 3.3, and we conclude that the bump is unconstrained. Thus, we decide to only consider slope deviations provided by  the $\delta$ parameter.

The stellar population models must  be attenuated before the IR emission can be added, since the total IR luminosity is defined as  the dust-absorbed luminosity of the stars.
The code allows us   to consider that the old stellar population is less affected by dust attenuation than the young one as recommended by several studies  \citep[e.g.][]{calzetti00, charlot00, panuzzo07}.  The primary input parameter is the dust attenuation in the V band of the young stellar population, $A_{\rm V,ySP}$.  A  reduction factor $f_{\rm att}$ of the dust attenuation for the old stellar population (relative to the young one) is also introduced as an input parameter. The global dust attenuation for the total (young and old) stellar population  is defined as the output parameter $A_{\rm V}$. The range of values adopted for    
$A_{\rm V,ySP}$ and  $f_{\rm att}$ are listed in Table \ref{table4}. Since we are particularly interested in dust attenuation in the UV range, we also define $A_{\rm FUV}$, the dust attenuation in the FUV filter.\\
To fit IR observations, CIGALE uses the semi-empirical one-parameter models of \citet{dale02}. These 64 models are parametrized by $\alpha$, which relates the dust mass to heating intensity.
 The parameter  $\alpha$ is directly related to  $R(60,100) = f_{60\mu m}$/$f_{100 \mu m}$, where $f_{60\mu m}$ and $f_{100\mu m}$ are  fluxes  in Jy at 60 and 100  $\mu$m, respectively.
  We choose $\alpha$ in the interval [1 ; 4]  to cover a large domain of activity.


\begin{figure}
\centering
 \includegraphics[width=9cm]{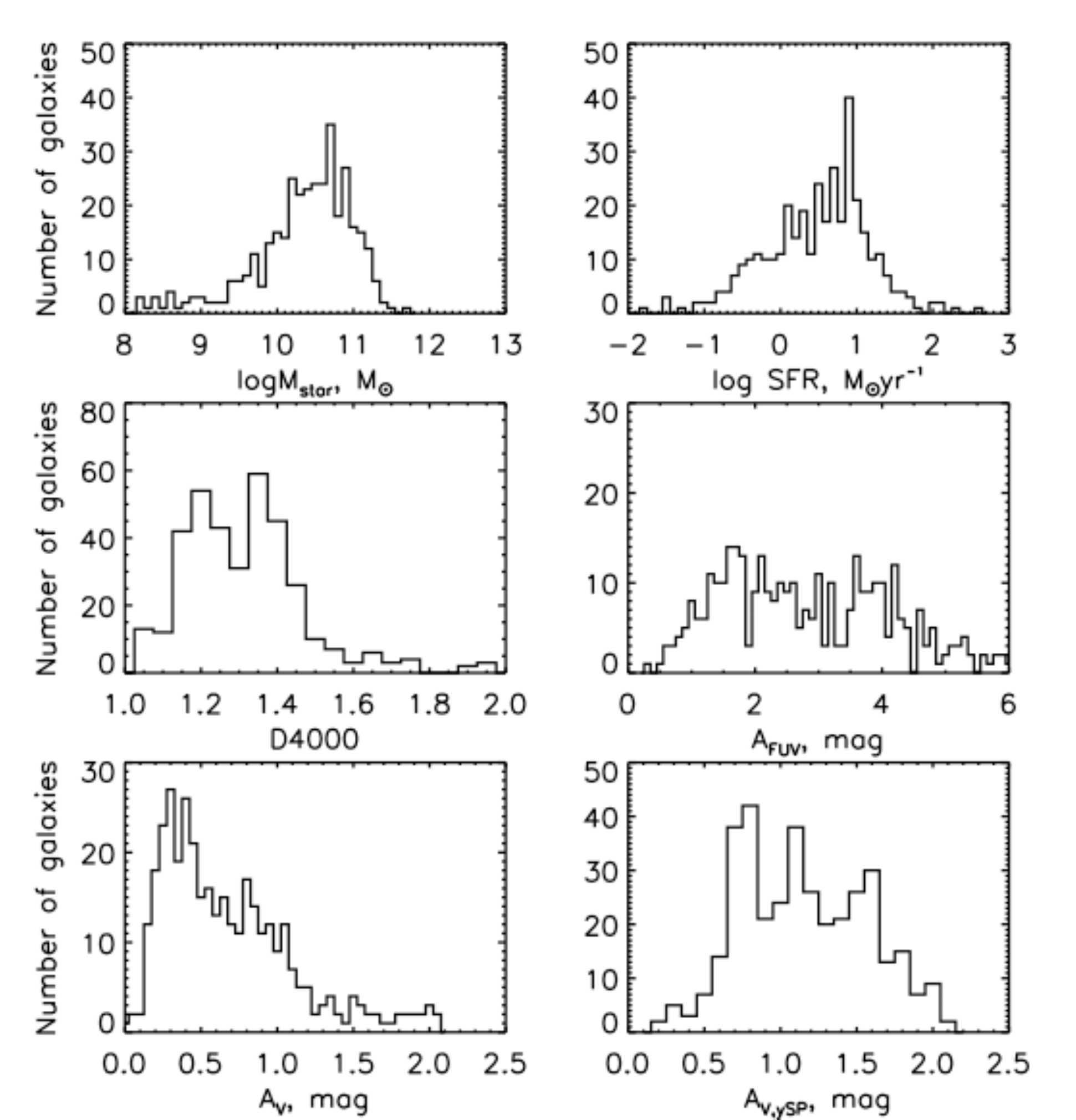}
 \includegraphics[width=9cm]{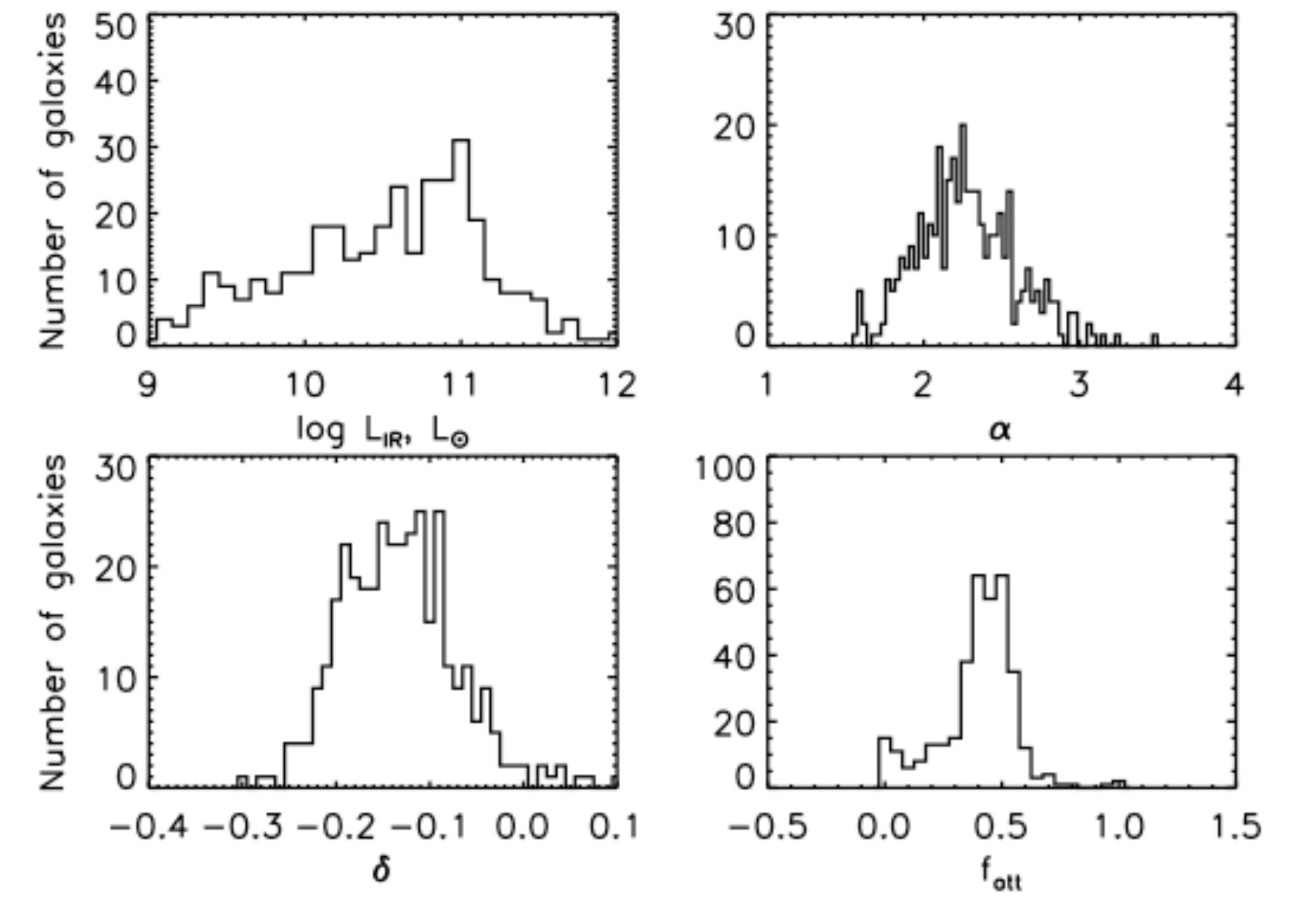}
    \caption{Distribution of the Bayesian estimates of  the output parameters discussed in this work: $M_{\rm star}$, SFR, D4000, $A_{\rm FUV}$, $A_{\rm V}$, $A_{\rm V, ySP}$, $L_{\rm IR}$, $\alpha$, $\delta$, and $f_{\rm att}$.}
     \label{HISTO}
\end{figure}

\subsection{Reliability and distribution of the output  parameters}
 A first check of the ability of the models  to reproduce the data is to compare their photometric distributions. The parameters we wish  to constrain are related to  the star formation history and dust attenuation characteristics that mainly affect the UV emission. Therefore, we choose to compare the distributions of the FUV-NUV and NUV-r colours, both being very sensitive to the star formation history and dust attenuation. One can see in  Fig~\ref{prior}  that the range of the modeled colours covers the range of the observed ones confirming that the stellar populations {\bf and the dust attenuation}  assumed in the models clearly  reproduce the  observed characteristics of the  sources . \\
  The reliability of  the parameter estimation is then tested by  applying the method developed  by \citet{giovannoli10}, which consists of  building a specific mock catalogue for each set of data to be fitted. The mock catalogue is  made of artificial galaxies for which we  know the flux in each photometric band of interest and  the exact values of the parameters used by CIGALE.
To construct this  catalogue, we follow three steps. The first step is to run the code on the data to choose the best-fit model for each object by means ot a simple {$\chi^2$} minimization method. We select the 28 galaxies detected at all IR wavelengths: 60, 100, 140, and 160 $\mu$m. In the second step, we add a randomly distributed error (which represents  typically 10$\%$ of the flux) to each flux of the best-fit models measured in the photometric bands of the dataset: 
we have a catalogue of artificial galaxies which are observed  in the same filters as the real galaxies and representative of them.  The representativeness of our mock catalogue is checked by comparing the colours (FUV-NUV and NUV-r) of the simulated and real galaxies (Fig~\ref{mock-colours}). The last step is to run the code on these simulated data and  compare the exact values of the parameters with the values estimated by the code.
Fig~\ref{mock}   compares  the output parameters   from the Bayesian analysis for the mock galaxies  considered in this work with the values estimated by the code.  Values of the linear Pearson correlation coefficient  (r) are also indicated in the plots.
The output parameters for the real  data are all Bayesian estimates. We find very good correlations for  $M_{\rm star}$, SFR, D4000, $A_{\rm FUV}$, $A_{\rm V}$, $A_{\rm V, ySP}$, $L_{\rm IR}$, and $\alpha$ with r $>$ 0.80. 
The mass fraction $f_{\rm ySP}$ is only poorly  estimated (r = 0.71) and the estimations of $\tau_1$, $t_2$, $\delta$, and $f_{\rm att}$ are much less satisfying,  with a correlation coefficient equal to 0.15, 0.52, 0.54, and 0.45  respectively. 
According to this  analysis which was also confirmed by  that of the Bayesian error distribution \citep{noll09b}, we conclude that  the reliable parameters are  $M_{\rm star}$, SFR, D4000, $A_{\rm FUV}$, $A_{\rm V}$, $A_{\rm V, ySP}$,  $L_{\rm IR}$, and $\alpha$. These parameters  are  used in the following, with some discussion about $\delta$ and $f_{\rm att}$.

The SED fitting and the Bayesian analysis are performed on the dataset with the input parameters listed in Table \ref{table4}. We restrict the analysis to objects for which minimum value of  the  reduced $\chi^2$ is lower than 10: 342  galaxies satisfy this criterium, the  reduced $\chi^2$ distribution for the entire sample is presented in Fig \ref{fits} (last panel). The discrepant cases are likely to correspond to misidentified objects and  {\it GALEX}   data that are inconsistent with the other ones. Given the low number of  these catastrophic cases ($\sim 6 \%$), we omit  them from  any further analysis. Some examples of best-fit $\chi^2$ models superposed on  the observed fluxes are given in Fig \ref{fits}.
 Fig.~\ref{HISTO} shows  the distributions for the parameters of interest estimated with the Bayesian analysis for the 342 objects: $M_{\rm star}$, SFR, D4000, $A_{\rm FUV}$, $A_{\rm V}$, $A_{\rm V, ySP}$,  $L_{\rm IR}$, $\alpha$,  $\delta$ and $f_{\rm att}$.\\
Galaxies  are  massive with a median value of $M_{\rm star}$ equal to  $10^{10.5}$M$_{\odot}$. 
We find that the SFR is between 0.1 and 100 $ M_{\odot} \mathrm{yr}^{-1}$ with very few values higher  than 30 $ M_{\odot} \mathrm{yr}^{-1}$ and a median  value of  4  $M_{\odot} \mathrm{yr}^{-1}$. The galaxies are moderately luminous ($<L_{\rm IR}> = 10.7 \pm 0.3 ~L_\odot$) as expected for a flux-limited sample of nearby galaxies. 
The D4000 distribution corresponds to that of actively star-forming galaxies in the nearby universe with a median value equal to 1.3  (\citet{kauffmann03a} and  discussion below). The parameter 
$A_{\rm V,ySP}$  spans the entire range of input values between 0.5 and 2.1 mag with a few objects above 1.25 mag. For most of the objects, $A_V$ $<$ 1 mag.   These output parameters  are  discussed in more detail in the following sections.

\subsection{Dust emission}
\subsubsection{Total infrared luminosities}
\begin{figure}
 \includegraphics[width=8cm]{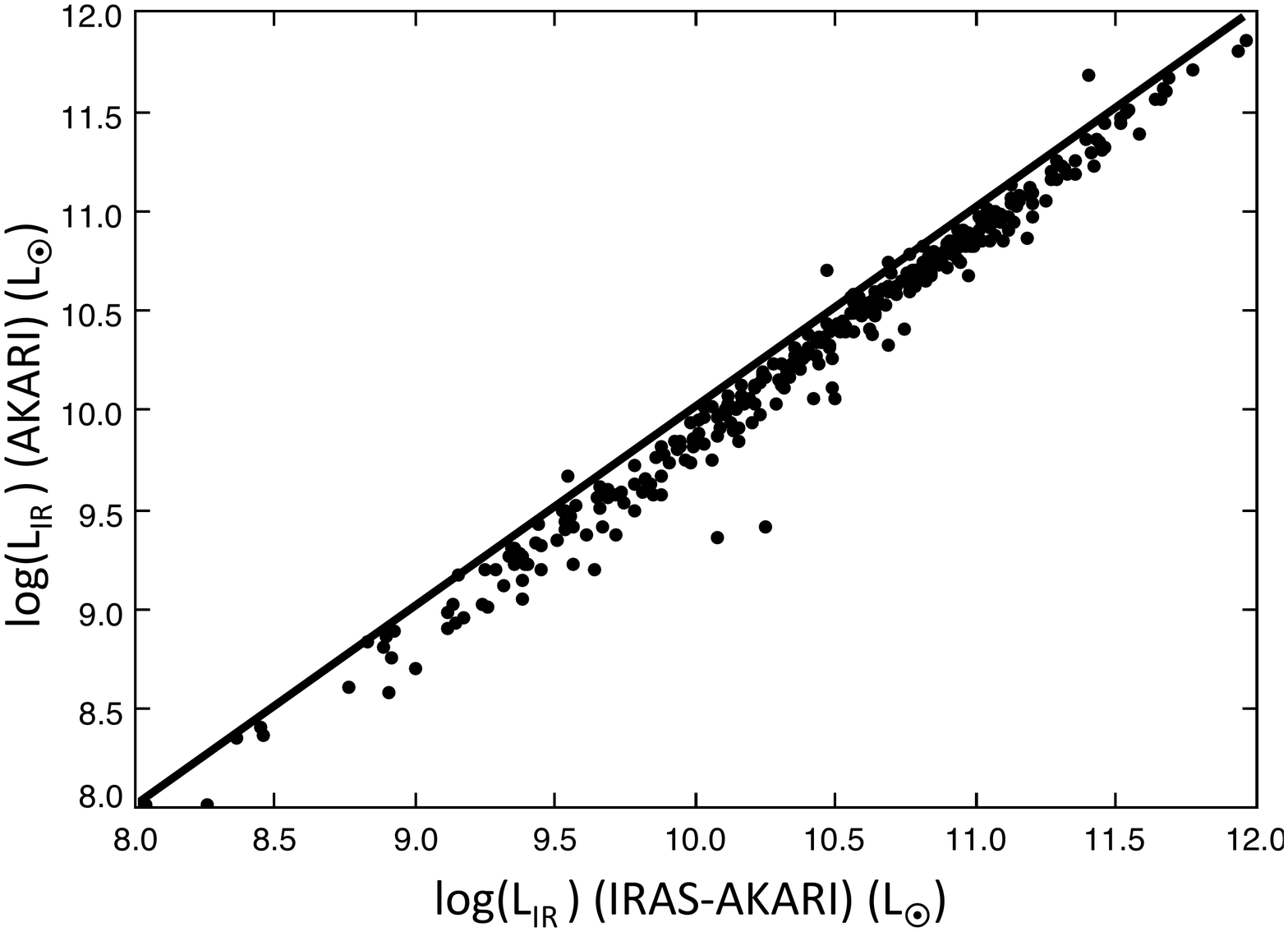}
  \caption{ Total IR emission $L_{\rm IR}$  obtained by running CIGALE with $IRAS$ and $AKARI$ data (x-axis) and with $AKARI$ data only (y-axis). The solid line correspond to equal values on both axes.}
      \label{calib-Lir}
\end{figure}

As discussed in section 2, {\it AKARI} measurements at 65 and 90 $\mu$m seem to be underestimated relative  to {\it IRAS} fluxes at similar wavelengths. We can compare the measure of the total IR luminosities $L_{\rm IR}$ obtained when considering the four {\it AKARI}-FIS bands or a combination of {\it IRAS} (at 60 and 100 $\mu$m) and {\it AKARI}-FIS (at 140 and 160 $\mu$m) data. We run  CIGALE  in both cases and the resulting values of $L_{\rm IR}$ are compared in Fig.~\ref{calib-Lir}. 
Both estimates correlate very well, the values obtained with the four $AKARI$ bands   being  systematically lower than those obtained by combining $AKARI$ and $IRAS$ data by  25$\%$. In the following, we only consider the combination of {\it IRAS} and {\it AKARI}  data. 

\subsubsection{Infrared SEDs}
\begin{figure}
 \includegraphics[width=8cm]{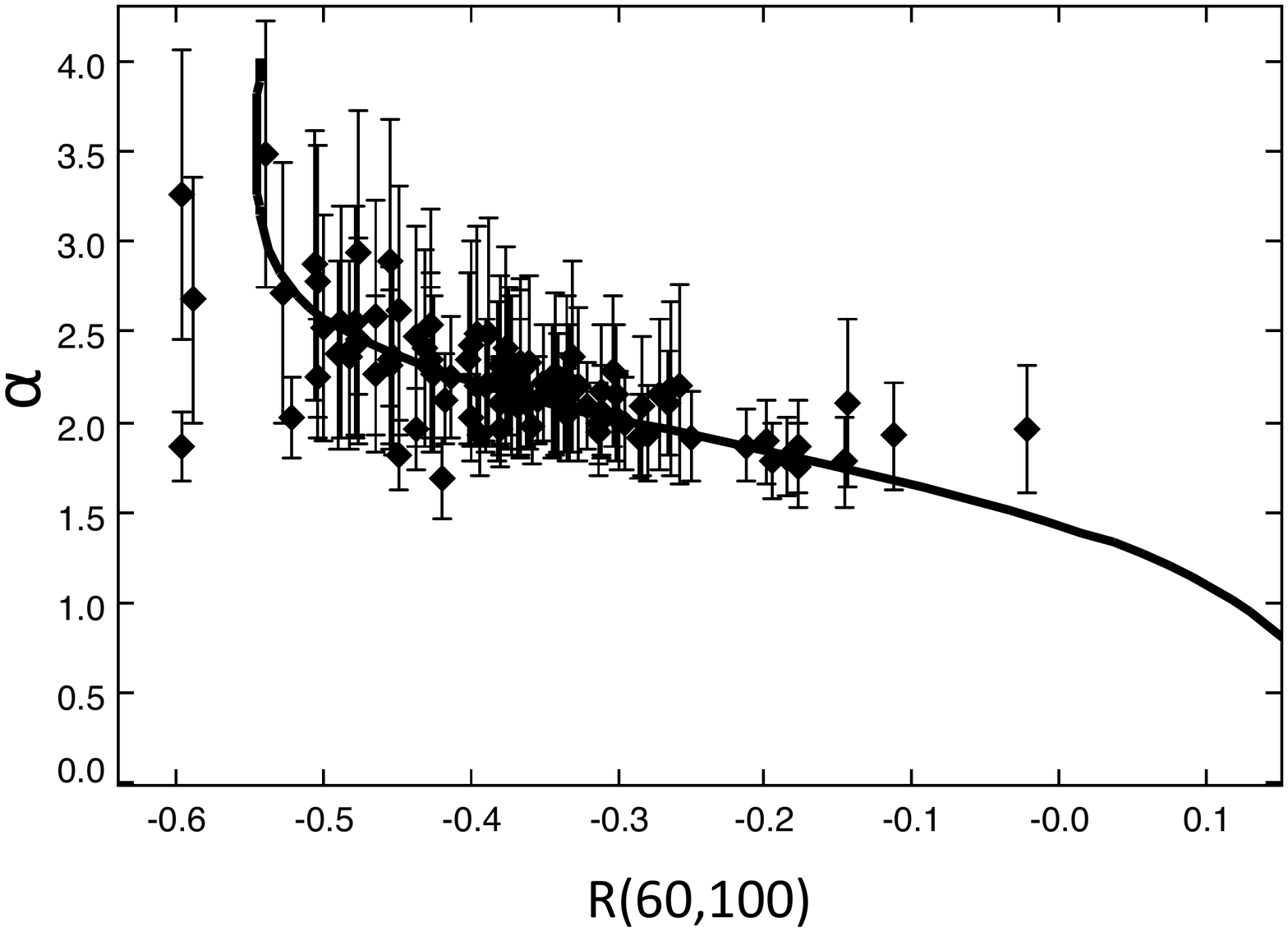}
  \caption{ $\alpha$ parameter estimated  with CIGALE as a function of the 60 to 100 $\mu$m flux ratio, R(60,100) for galaxies observed at all far-IR wavelengths, up to 160 $\mu$m.  Vertical error bars represent the standard deviation in the determination of $\alpha$ given by the Bayesian analysis. The relation between $\alpha$ and R(60,100) for the DH02 models is plotted as a solid line.}
      \label{alpha}
\end{figure}

The IR dust emission is fitted using the \citet{dale02} library (hereafter DH02). These models are parametrized by  a value of $\alpha$ that is directly related to R(60,100) the ratio of the fluxes at 60  to  100 $\mu$m observed by {\it IRAS}.  In Fig.~\ref{alpha}, the values of $\alpha$ obtained for galaxies observed at 60, 100, 140, and 160 $\mu$m are plotted against the ratio of the {\it IRAS} fluxes at 60  to  100 $\mu$m, R(60,100), and compared to the relation between $\alpha$ and R(60,100)  from the DH02 models. A very good agreement is found between $\alpha$ values found with CIGALE and the values expected from R(60,100). The SED library of  DH02  is  built to clearly  reproduce   the   far-IR (longward 100 $\mu$m) and  sub-mm (at 850 $\mu$m) emission of local galaxies. Therefore,    good agreement is expected when data longer than 100 $\mu$m are introduced  and   the SEDs are likely to be better constrained when both  data at 140 and 160 $\mu$m are added.  When galaxies are only observed at 60, 100, and 140 $\mu$m,   $\alpha$  is found to be slightly larger ($\rm <\alpha(measured)-\alpha(predicted)> \simeq$ 0.2) than predicted from R(60,100) and the DH02 relation: we can tentatively explain this trend by the  high values of $\alpha$  being  likely  to be  ill-constrained without data at 160 $\mu$m leading to a degeneration of the probability distribution and then to   a slight  overestimation of $\alpha$.  In all cases the difference in the $\alpha$ determinations remains  lower than the uncertainty given by the Bayesian analysis.
\subsubsection{Constraints on the dust attenuation curve}
The presence of a bump at 220 nm in the attenuation curve of external galaxies is  still controversial. The best way to search for  the presence of this bump  is to work with UV spectra. In   pioneering studies,   \citet{calzetti94} found no bump in local starburst galaxies, \citet{burgarella05} found some hints of  a bump in one  local Luminous IR Galaxy observed with {\it GALEX}, and,  at higher z \citet{noll09a} found  evidence of  a bump in  star-forming galaxies with $1<z<2.5$.  Although  the NUV band of {\it GALEX}   lies on the bump itself \citep{conroy10},  our SED fitting is  unable to  reliably ascertain whether  a bump is present,  as mentioned in section 3. We can say that the slope of the UV attenuation curve (parameter $\delta$) is more reliable although the uncertainty remains large. The number of free parameters acting on the UV emission (star formation history, various parameters of dust attenuation) are at the origin of these poor determinations as discussed by  \citet{noll09b}.

\begin{figure}
 \includegraphics[width=8cm]{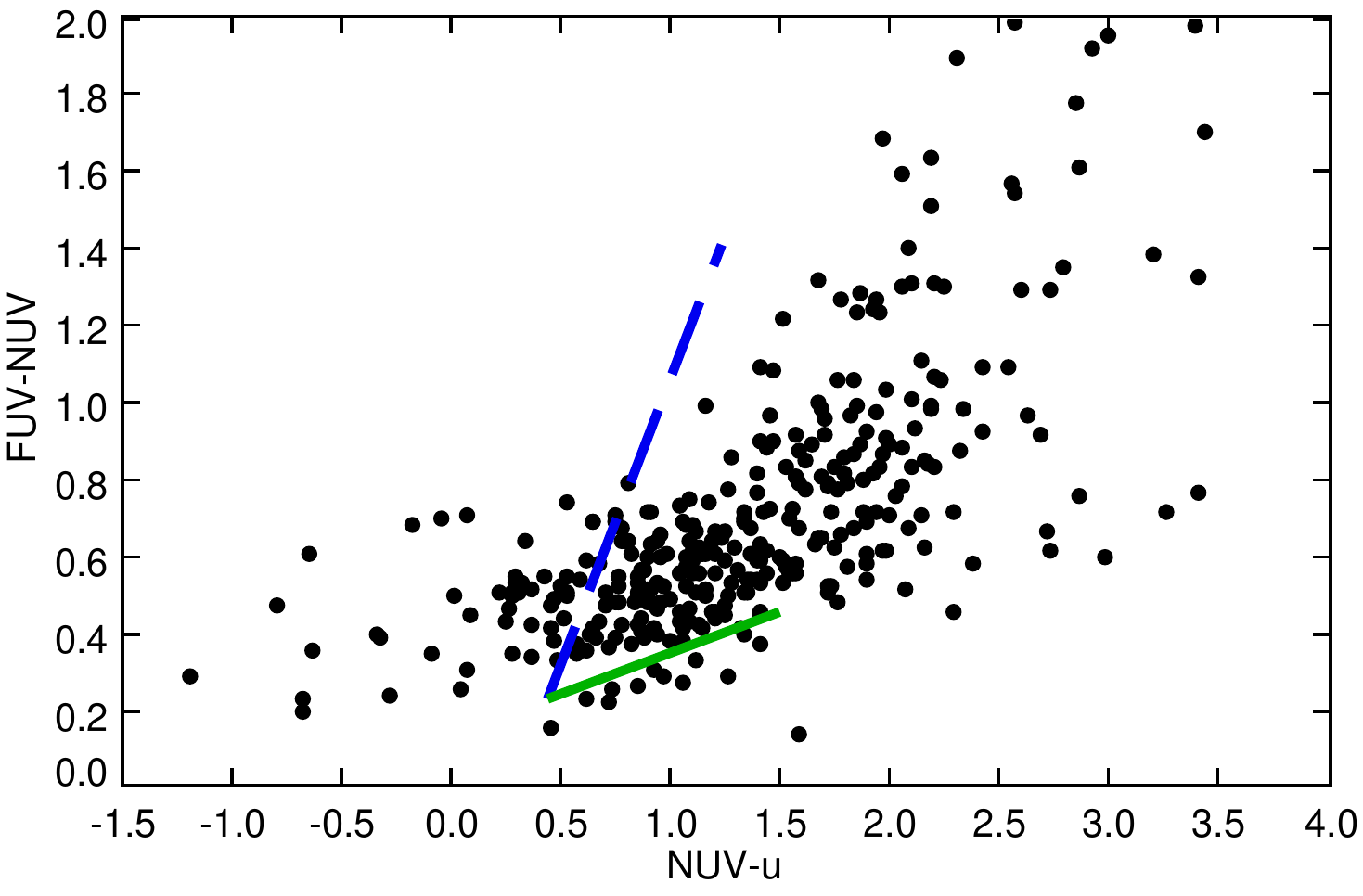}
  \caption{$ NUV-u$ versus $FUV-NUV$ observed colours (black points) compared to \citet{conroy10} models (lines): the dashed blue line corresponds to a dust attenuation curve with no bump and the green solid line to a Milky Way bump strength.}
      \label{colorplot}
\end{figure}

\citet{conroy10} adopted  a more empirical approach  and developed a diagnostic based on a colour analysis that  clearly distinguishes models with or without a bump. 
 The comparison of the $FUV-NUV$ and $NUV-u$  colours is sensitive to the presence of the bump. Their assumption of a constant star formation rate over 13.7 Gyr  considerably  reduces the number of  free parameters. They also  assumed a  steep underlying curve (by adopting a low R$_V$)  and found that their galaxies have $FUV-NUV$ and $NUV-u$ colours compatible  with the presence of a bump with an amplitude slightly lower than for the Milky Way extinction curve. 
 In Fig~\ref{colorplot}, we   report  the colours of our galaxies with the predictions of \citet{conroy10} for no bump or a bump with an amplitude similar to that of the Milky Way. Our galaxies span a larger range of colours than those explored by \citet{conroy10} but they clearly lie between the two predicted lines, suggesting that   a bump is present that has a lower  amplitude than for  the MW. In all cases, the analysis of the full SED of these galaxies allowing for variations in the star formation history and dust attenuation scenarios cannot reach a reliable conclusion about  the presence of a bump.  \\
 The parameter $\delta$  quantifies the variations in  the slope of the attenuation law, the reference being the starburst reddening curve of \citet{calzetti00}. For the measurement of $<\delta> = -0.13 \pm 0.07$,  the slopes of  the attenuation law are found to be   slightly steeper than   the Calzetti law  but $\delta$ is not well constrained with the available data (Fig~\ref{mock}). This  illustrates the difficulty in constraining the wavelength dependence of the attenuation curve with broad-band data only. As explained in \citet{noll09b}, $\delta \not= 0$ corresponds to  a variation in  $R_V$ originally equal to 4.05 for the Calzetti et al. law: $<\delta> = -0.13  $ corresponds to $R_{\rm V} = 3.6$.\\
 
\subsection{Age and mass of the stellar component}
\begin{figure} 
\includegraphics[width=9cm]{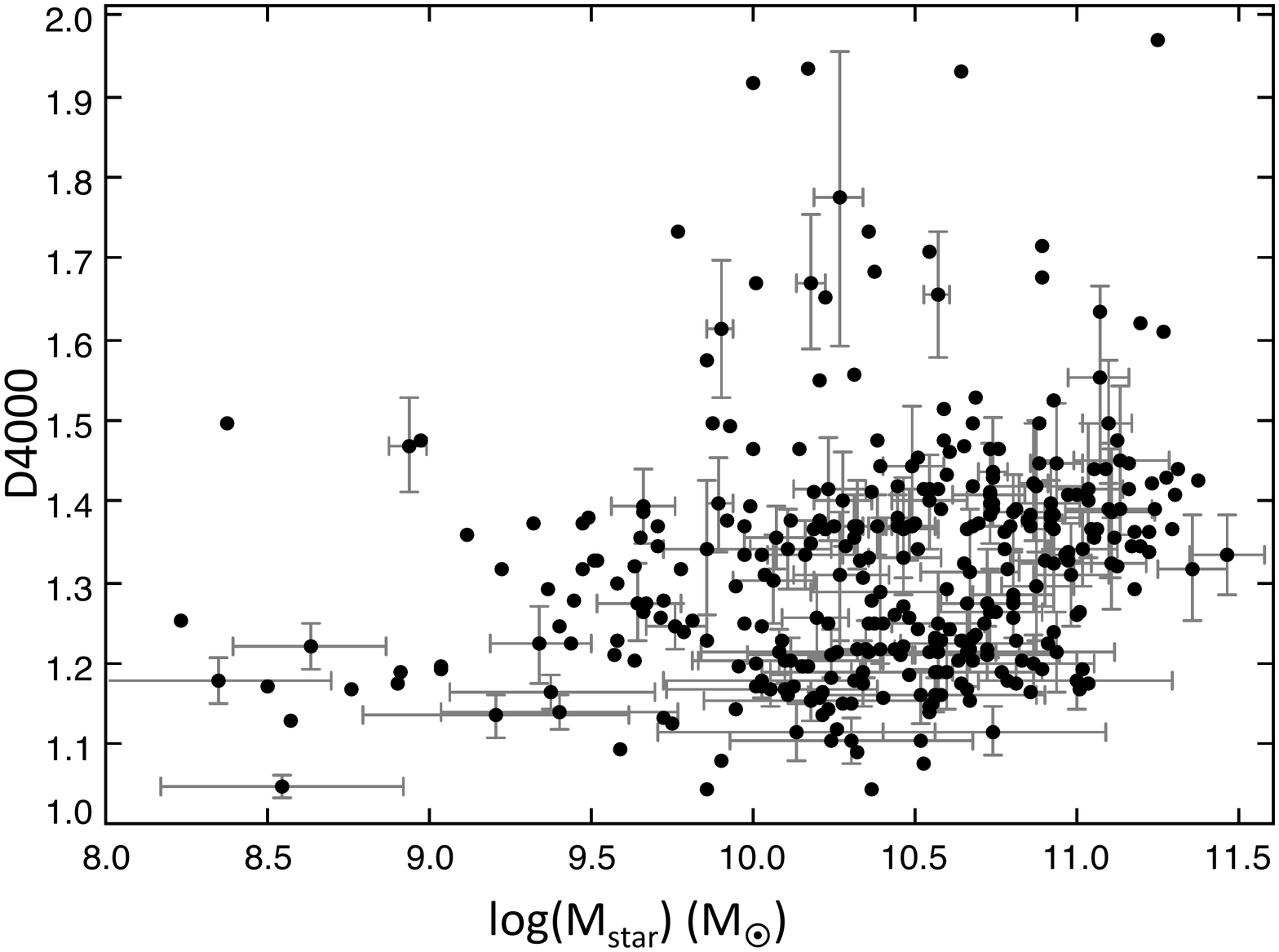}
  \caption{Strength of the dust free 4000 $\AA$ break, D4000, plotted against the stellar mass of the galaxies: both values are output parameters of the SED fitting code, the standard deviations given by the Bayesian analysis are plotted as error bars for 1/4 of the sample.}
      \label{age-mass}
\end{figure}
\begin{figure} 
\includegraphics[width=9cm]{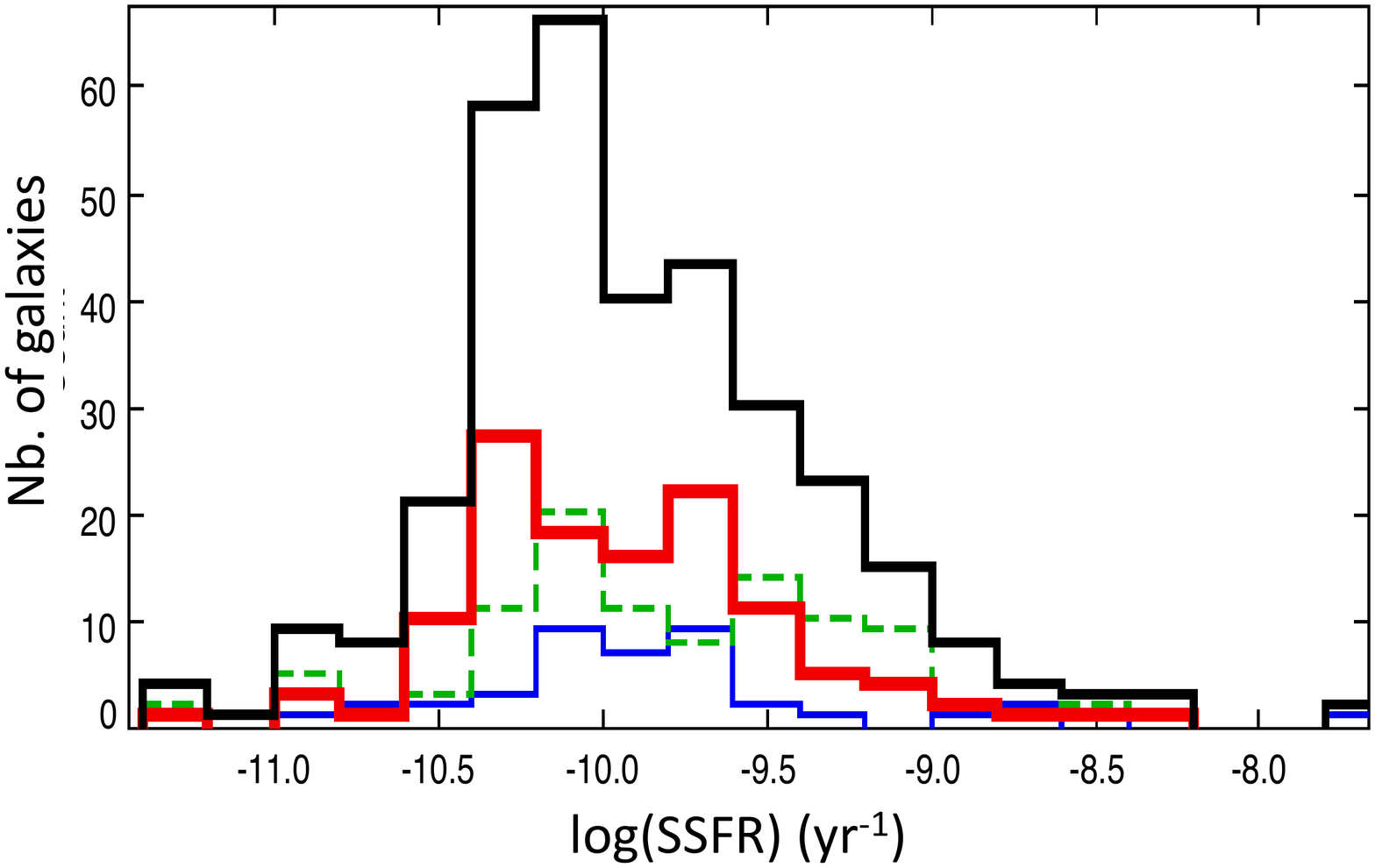}
  \caption{Histogram of the specific star formation rate (SSFR) for the whole sample (heavy solid black line), and per mass bin (for the mass ranges  $9.5< \log(M_{\rm star})<10$ (thin solid blue line),  $10< \log(M_{\rm star}<10.5)$ (dotted green line), and  $10.5< \log(M_{\rm star})<11$ (heavy solid red line). }
      \label{ssfr}
\end{figure}

The  4000 $\AA$  break (hereafter D4000) provides  information on the age of the stellar populations. It is small for young stellar populations and large for old galaxies \citep{kauffmann03a}.  The D4000 definition adopted here is that of \citet{balogh99}. It is expected to be insensitive to dust attenuation. Our models  have a resolution that is sufficient to calculate the D4000 break \citep{maraston05}, the calculation being based  on the un-reddened spectra and thus an intrinsic, dust-free D4000 measurement.   D4000 is quite sensitive to metallicity and its variation with the age of the stellar population is also strongly dependent on the star formation history \citep{maraston05}. Its smooth and continuous increase when stars become older makes it a useful indicator of the age of the  predominant stellar population. Moreover, its extensive use in  analyses of  SDSS data sets \citep[e.g.][]{kauffmann03b} has provided  references for the distribution of the D4000 in the nearby universe. 
In Fig~\ref{age-mass}, we   report the D4000 values as a function of the stellar mass, both  quantities being  outputs of  CIGALE.  The value of D4000 (i.e. the age of the stellar populations) increases with the stellar mass, as also found by \citet{kauffmann03b}. Most  objects have a low D4000 (D4000$<$1.5), even for  massive galaxies. The situation  is quite different for the SDSS galaxies that  exhibit a bimodal D4000 distribution (roughly separated at D4000 = 1.5), the more massive galaxies ($M_{\rm star} > 10^{11}~M_\odot$) having D4000 $> 1.5$. Our galaxies have  D4000 values characteristic of   galaxies with $10^9<M_{\rm star} <  10^{10} ~M_\odot$ in the SDSS. \\
Another interesting parameter is the specific star formation rate SSFR = SFR/$M_{\rm star}$ which is a measure of the present star formation activity relative  to the past one and gives a  rough indication of the star formation history in galaxies. The histograms of SSFR for the whole sample and different mass bins are plotted in Fig~\ref{ssfr}. As expected from their D4000 and stellar mass distributions, galaxies exhibit a unimodal distribution of SSFR with a tail towards high SSFR, the  average value for the whole sample being  $10^{-10}$ yr$^{-1}$. This  distribution is similar to that found for galaxies with stellar mass  between $10^9$ and $10^{10}$ $M_\odot$ in the SDSS \citep{brinchmann04}. No obvious trend is found with stellar mass, in contrast  to the findings of   \citet{brinchmann04}  in the SDSS. \\
 We can therefore conclude that our  selection, which is  mainly based on a detection at 140 $\mu$m with {\it AKARI}, is biased towards    massive galaxies ($<\log(M_{\rm star})> = 10.4\pm 0.6 [M_\odot]$) that are quite active in star formation relative   to the  SDSS galaxy sample: the SSFR and D4000  found for our sample are representative of less massive galaxies when the galaxies  are selected using visible data.

\section{ Dust attenuation}
\begin{figure}
  \includegraphics[width=9cm]{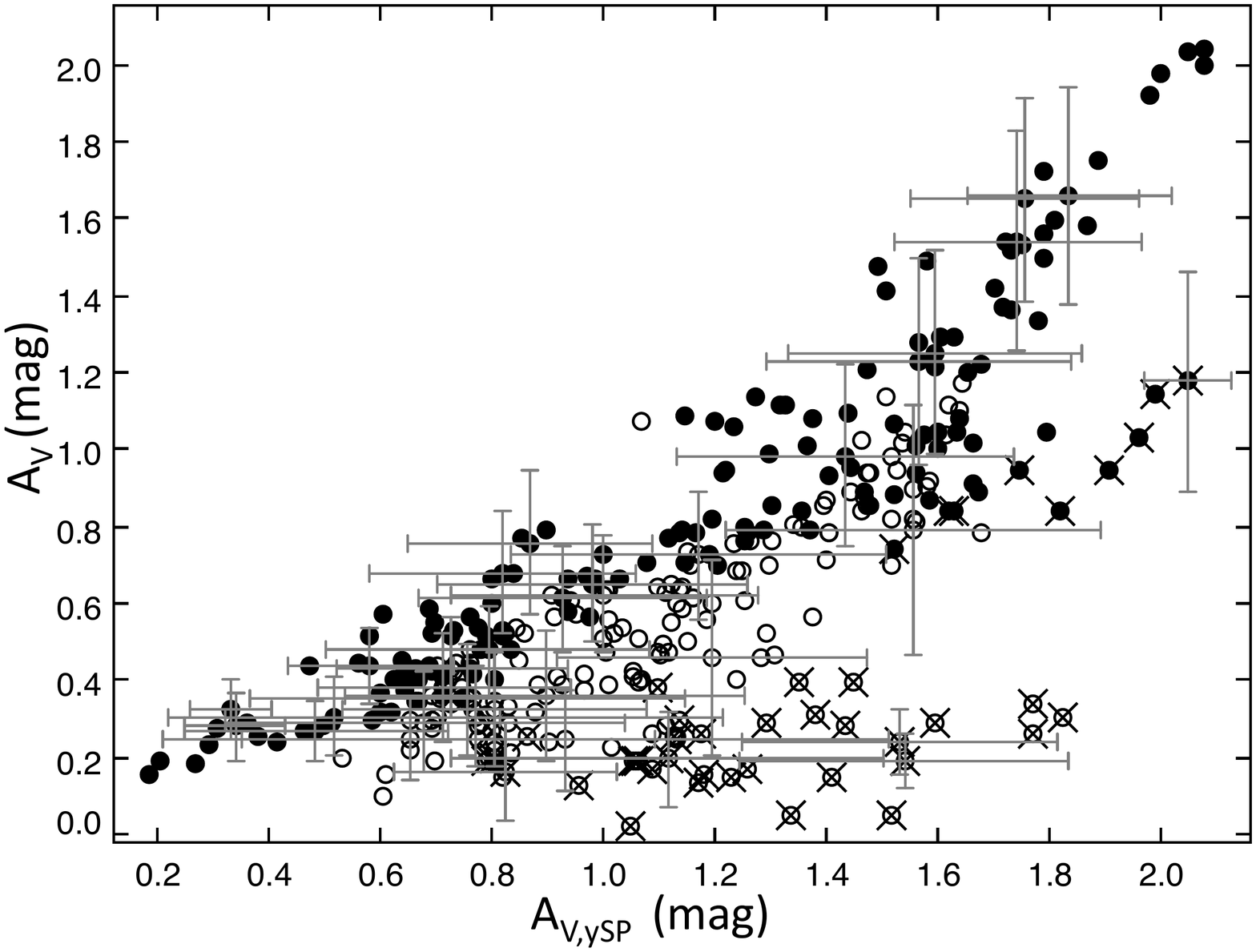}
  \caption{ V-band attenuation, $A_{\rm V}$, plotted against V-band attenuation of the young stellar population alone,  $A_{\rm V,ySP}$.  Filled symbols are for galaxies with   $\rm D4000< 1.3$  and  empty ones for objects with $\rm D4000> 1.3$. Standard deviations are plotted as error bars for 1/4 of the sample. Crosses are over-plotted on symbols corresponding to galaxies with $f_{\rm att} < 0.2 $.}
      \label{Av-comp}
\end{figure}
\begin{figure}
 \includegraphics[width=9cm]{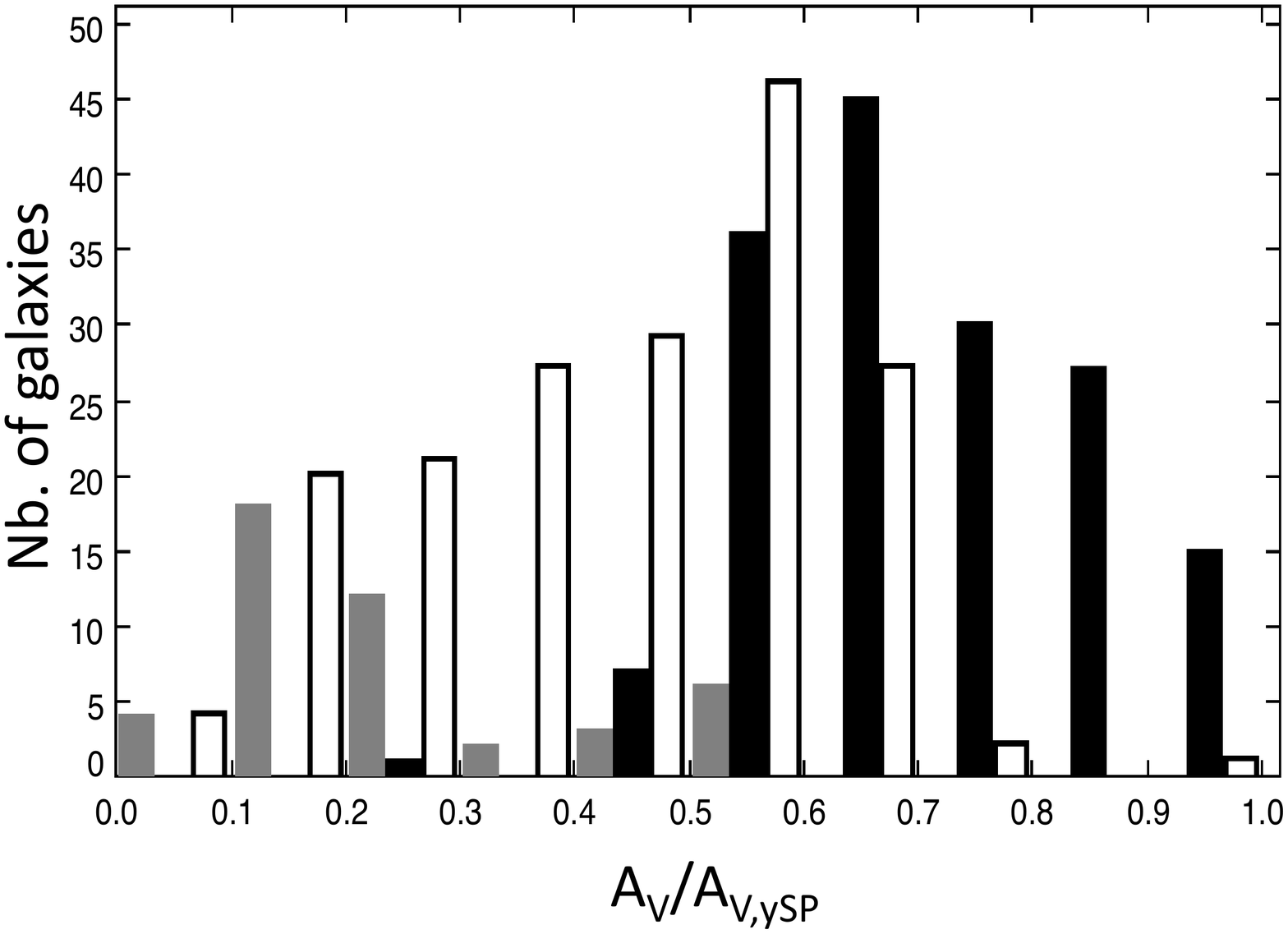}
  \caption{ Distribution of the ratio of V-band attenuation over V-band attenuation of the young stellar population alone,  $A_{\rm V}/A_{\rm V,ySP}$. The empty histogram corresponds to galaxies with   $\rm D4000> 1.3$, the  black filled  histogram  to galaxies with  $D_{\rm 4000}< 1.3$ and the grey filled histogram to  galaxies with $f_{\rm att} < 0.2. $}
      \label{Av-hist}
\end{figure}

\subsection{ Time-dependent dust attenuation}
Young stellar populations  may  be more affected by dust attenuation than  older ones \citep{calzetti00, charlot00, panuzzo07}. CIGALE allows us  to  separately vary dust attenuation for the two stellar populations considered for the star formation history by adjusting  the parameter $f_{\rm att}$, the   reduction factor  applied to the dust attenuation in the V band of the old population (expressed in magnitudes).  Although $f_{\rm att}$ is poorly constrained, the global dust attenuation $A_{\rm V}$  (for both old and young  stellar populations) and that found for the young stellar population alone  $A_{\rm V,ySP}$ are clearly determined.  We consider here D4000 as an  estimate of the luminosity-weighted age of the  stellar population and  we divide the sample into two sub-samples corresponding to  D4000 lower or larger than 1.3. In Fig~\ref{Av-comp}, the   total V-band  attenuation $A_{\rm V}$ is plotted against the   V band attenuation of the young stellar population only,  $A_{\rm V,ySP}$ . The quantities are  correlated and  their relation  depends on  the stellar population age.
 The galaxies with D4000 $>$ 1.3 for which $f_{\rm att}$ is found to be  lower than  $0.2$ exhibit a very low $A_{\rm V}$: the old stellar population contributes substantially to the stellar emission and  undergoes  dust attenuation that is much lower than the one affecting the young component. This is indicative of  a low value for the global attenuation in the V band relative to that applied  to the young stellar population. \\
 Histograms of $A_{\rm V}/ A_{\rm V,ySP}$ are plotted in Fig~\ref{Av-hist}. As expected,  the ratio is found to be  lower for galaxies with $\rm D4000> 1.3$ : $<A_{\rm V}/A_{\rm V,ySP}>= 0.43 \pm 0.17$ ($0.49 \pm 0.13$ for sources with  $f_{\rm att} > 0.2$) than for galaxies with $\rm D4000< 1.3$:  $<A_{\rm V}/A_{\rm V,ySP}>= 0.70 \pm 0.14 $.\\ 
\citet{calzetti00} found a factor of 0.44 between the color excess of the emission lines (related to the very young stellar populations) and the underlying continuum (older stellar populations). \citet{charlot00} recommend  reducing  by a factor three the effective optical depth for stars older than $10^7$ years. In the same way  \citet{panuzzo07} defined different age-dependent scale-heights for the stellar population, using  again a typical timescale of $\sim 10^7$ years to define young stars. In the present work,  the young and old stellar populations are defined differently from these previous studies:  the young stellar population does not represent stars younger than $10^7$years but the stellar population  from a second episode of star formation and the parameter D4000  measures a mean age of the global stellar populations dominating the optical emission. Although the age of the young stellar population alone ($t_2$) is ill-constrained (cf. section 3),  the mean value of the output parameter $t_2$  is $\sim 0.25-0.3$ Gyr, i.e. older than $10^7$years. Moreover, we compare the attenuation of the global (old$+$young) stellar population $A_{\rm V}$, which also depends on the star formation history,  to that of the young stellar population alone, $A_{\rm V,ySP}$. We cannot therefore directly compare   with  the difference of color excess E(B-V) found by \citet{calzetti00} between the nebular gas emission lines and the underlying stellar continuum,  which correspond to  more extreme  ages for the stellar populations (a few $10^6$ years for the emission lines and a  few Gyr for the underlying continuum of the Balmer lines).  The   recipe of \citet{charlot00}, which consists of reducing by a factor three the effective absorption after 10$^7$ years, is  related to our parameter $f_{\rm att}$, which is  defined as the ratio of $A_V$ for the old to young stellar populations. As discussed  in section 3, $f_{\rm att}$ is  ill-constrained  for individual cases. Nevertheless, $f_{\rm att}$ = 1 is chosen as the best-fit model  when running CIGALE ($\chi^2$ analysis) for only 4 $\%$ of the galaxies and  the Bayesian analysis gives $f_{\rm att} > 0.7$ for only seven  sources (cf. Fig.~\ref{HISTO}). The mean  value of $f_{\rm att}$ for the whole sample is 0.39$\pm$ 0.15, below the mean of the input values (equal to 0.5, cf. Table~\ref{table4}),  which, for an ill-constrained parameter, suggests a true mean lower than the return value of 0.39.    Hence we can conclude that   $f_{\rm att}< 0.5 $ as already found by \citet{noll09b} for SINGS galaxies.  An attenuation factor lower than 0.5 as well as  a global dust attenuation  ($A_{\rm V}$) lower than  that of the young stellar population alone ( $A_{\rm V,ySP}$)  provide strong support  for an age-dependent process for dust attenuation in galaxies. 
Given the uncertainties in the exact determination of $f_{\rm att}$ and  the different timescales adopted to define young stellar populations, our results are consistent with those of \citet{calzetti00} and \citet{charlot00} and an age-dependent dust attenuation has to be assumed when performing SED fitting.

    \subsection{ $A_{\rm FUV}$ as a function of  $L_{\rm IR}/L_{\rm FUV}$}
\begin{figure}
 \includegraphics[width=9cm]{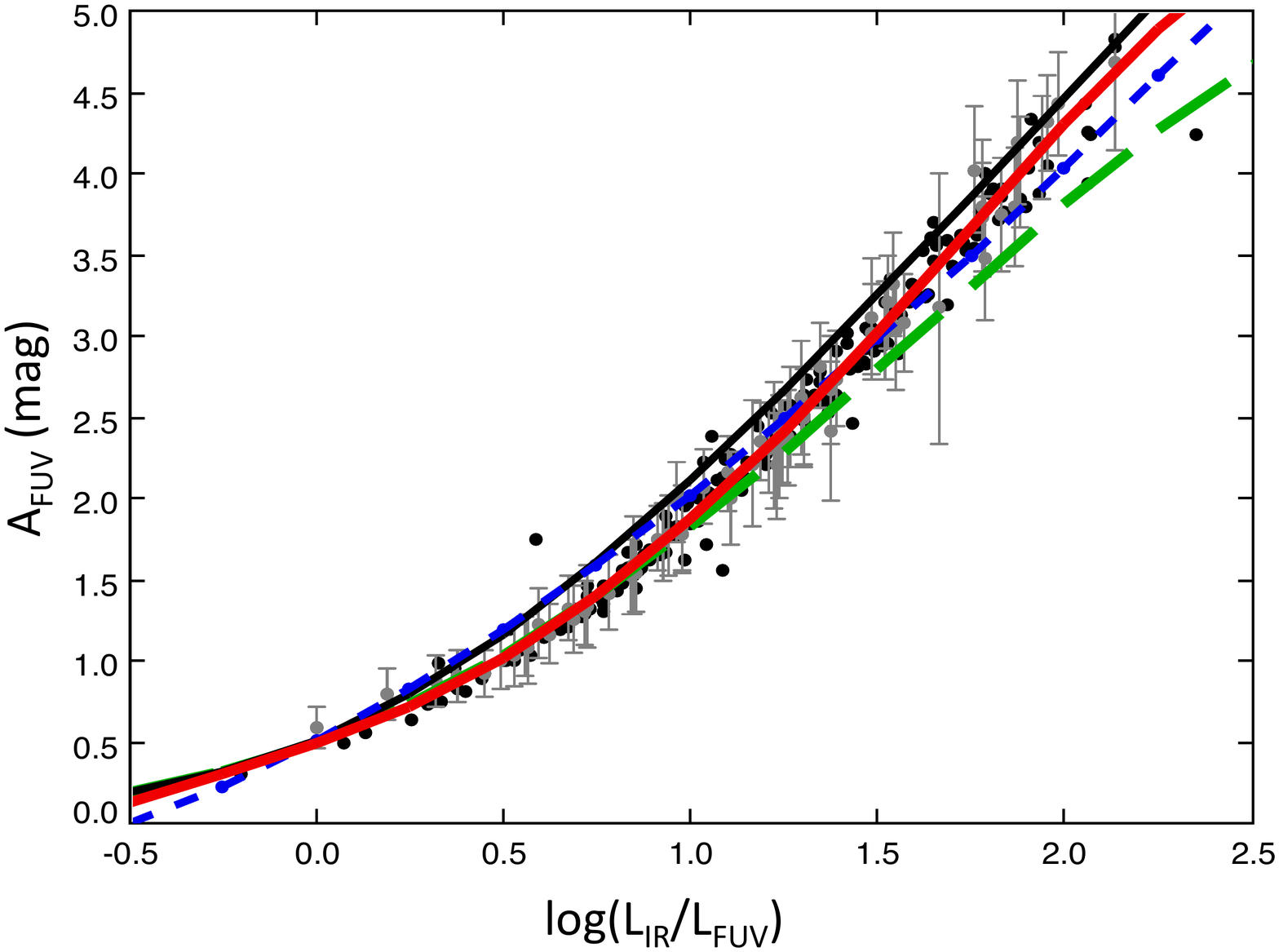}
  \caption{ $A_{\rm FUV}$ plotted as a function of $ L_{\rm IR}/L_{\rm FUV}$,  the result of the polynomial regression being  plotted as a  red  thick  solid  line. Models from \citet{Meurer99} (black thin solid line), \citet{buat05} (blue dotted line),  and  \citet{cortese08} with $\tau$= 6.2 Gyr (green dashed line) are overplotted. Standard errors in  $A_{\rm FUV}$ are plotted for 1/4 of the objects.}
      \label{AFUV}
     \end{figure}

    The $L_{\rm IR}/L_{\rm FUV}$ ratio\footnote{$L_{\rm FUV} = \nu. L_\nu$ at 153 nm  expressed in solar units as $L_{\rm IR}$} is known to be a very robust estimator of  $A_{\rm FUV}$ as long as galaxies form stars efficiently. Several calibrations can be found, the main difference coming from the star formation history adopted \citep[e.g.][]{buat05,cortese08}. In quiescent systems, a substantial part of  dust heating  is due to photons   emitted by  old stars which do not  emit primarily in UV. In this case, the relations between $L_{\rm IR}/L_{\rm FUV}$ and $A_{\rm FUV}$ derived for an active star formation overestimate dust attenuation in the UV range.
    CIGALE gives the global attenuation in the FUV (combination of dust attenuation of young and old stellar populations), which is plotted in Fig.~\ref{AFUV}  as a function of $ L_{\rm IR}/L_{\rm FUV}$. A very tight relation is found, as  expected since our galaxies are found to constitute a homogeneous population of star-forming galaxies in terms of  SSFR and D4000. Applying  a polynomial regression on our data yields:
    $$ A_{\rm FUV} = 0.483+0.812~y+0.373~y^2+0.299~y^3-0.106~y^4$$, where  $y =\log(L_{\rm IR}/L_{\rm FUV})$.\\
     We can compare these estimates to  other  $A_{\rm FUV}- L_{\rm IR}/L_{\rm FUV}$ relations from the literature, obtained with different  scenaros for the  dust attenuation and star formation rates. \citet{gordon00} showed that the relation between  $A_{\rm FUV}$ and $ L_{\rm IR}/L_{\rm FUV}$ is insensitive to  dust attenuation configurations. Further studies identified a stronger  sensitivity to the star formation history.  \citet{Meurer99} assumed a young stellar population only, for constant star formation rate over 1 to 100 Myr,  to derive a relation between $A_{\rm FUV}$ and $IRX$ ($IRX$ is defined  as  $L_{\rm FIR}/L_{\rm 1600}$, the ratio of the far-IR luminosity between 40 and 120 $\mu$m to  the luminosity at 1600 $\AA$). \citet{cortese08} explored more realistic models for normal galaxies with exponentially decreasing SFRs and  various $\tau$,  \citet{buat05} obtained a mean relation  for  a large number of star formation scenarios, from short bursts to constant star formation rates. These different relations are reported in Fig.~\ref{AFUV}.    All of them are obtained by assuming  a single stellar/dust configuration, without any time dependence of the attenuation whereas a different attenuation for the old and young stellar population is introduced in CIGALE.  The  relation of \citet{Meurer99} systematically overestimates the dust attenuation we find in this work. Our results are more  compatible with the relation of  \citet{buat05} at least for  galaxies with dust  attenuations larger than 2 mag. The   \citet{cortese08}  relation with  $\tau = 6.2$ Gyr provides a good  fit for sources with intermediate attenuation (between 1 and 2 mag).  In all cases the variations in  these estimates remain modest there being a difference of  at most 0.3 mag between our polynomial regression and the relations of \citet{Meurer99} and \citet{buat05} that  reaches 0.7 mag with the \citet{cortese08} calibration for large dust attenuations.
   
\subsection{$A_{\rm FUV}$ and UV-optical colours}
     
\begin{figure}
 \includegraphics[width=9cm]{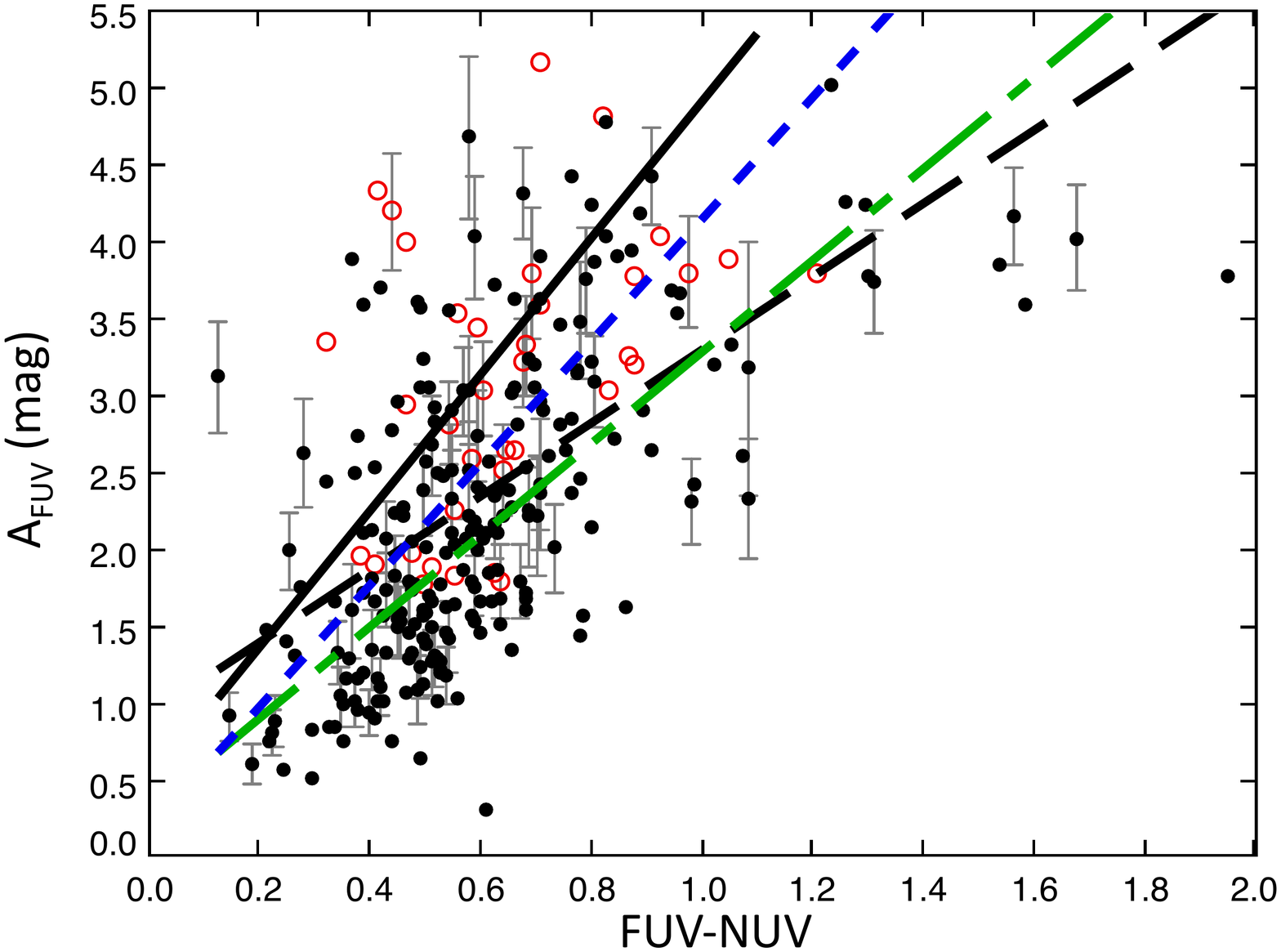}
  \caption{$A_{\rm FUV}$ plotted against $FUV-NUV$. The sample is restricted to 267 galaxies with an error in  $FUV-NUV$ lower than 0.2 mag, and the linear regression is plotted as a  dashed black line. Galaxies with $L_{\rm IR} > 10^{11}~L_\odot$   are plotted as empty red circles. Relations from previous studies are overplotted: \citet{salim07} (green dot-dashed line), \citet{seibert05} (blue dotted line), and \citet{Meurer99} (black solid line). Standard errors in  $A_{\rm FUV}$ are plotted for 1/4 of the objects.}
      \label{AFUV-colour}
     \end{figure}
\begin{figure}
 \includegraphics[width=9cm]{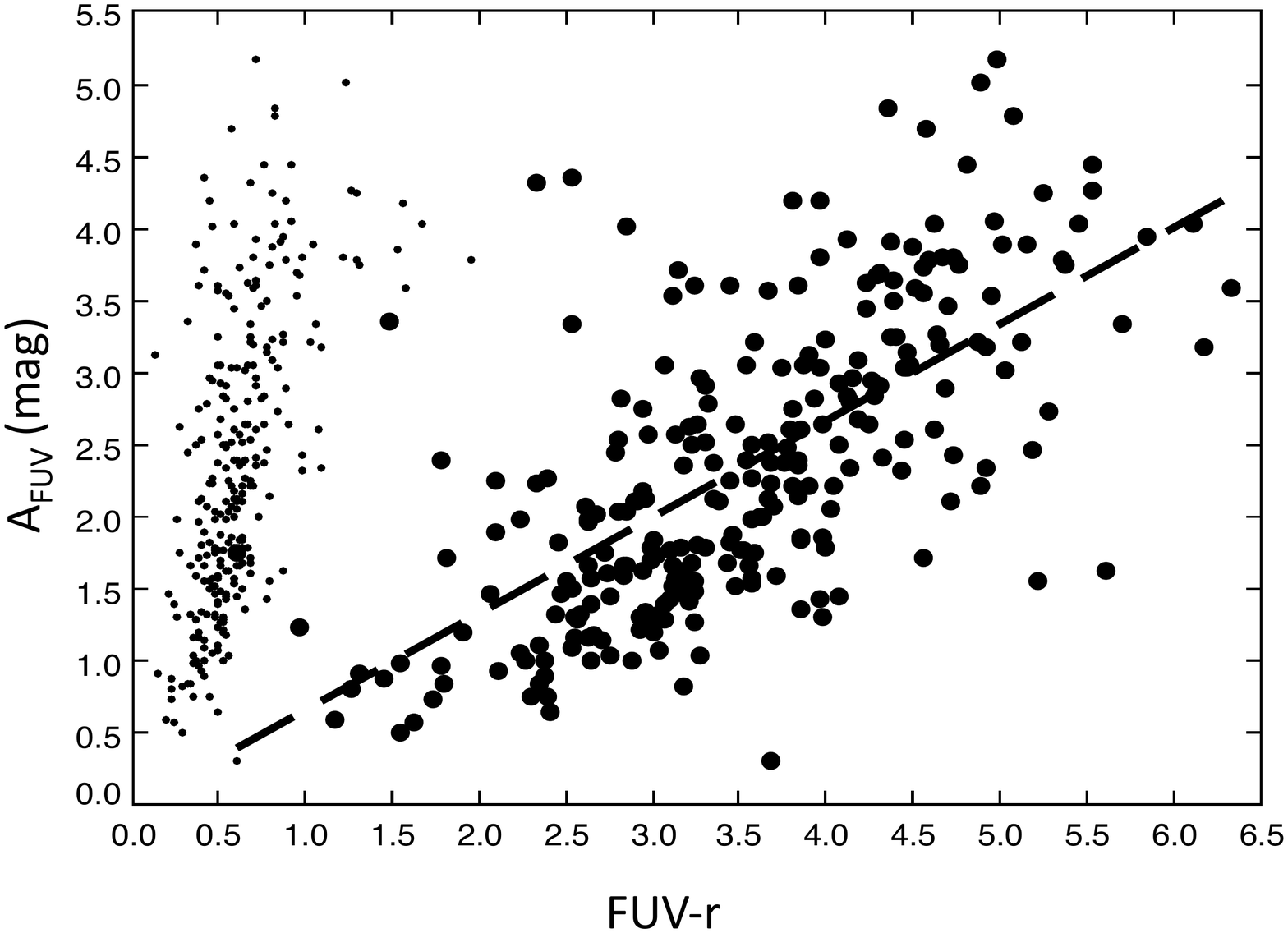}
  \caption{$A_{\rm FUV}$ plotted against $FUV-r$. The sample is restricted to 265 galaxies  with an error in $FUV-r$ that is smaller  than 0.2 mag, the linear regression  is plotted as  a dashed line.  The  small dots correspond to $FUV-NUV$ reported on the x-axis (similar to Fig.~\ref{AFUV-colour}) and are overplotted here for comparison (see text for details).}
      \label{AFUV-colour2}
     \end{figure}
Dust attenuation diagnostics based on UV-optical data alone can be  very useful for applyung a  correction  for dust attenuation to  the emission of galaxies not observed in far IR.   \citet{Meurer99} found a relation between the slope of the rest-frame UV continuum $\beta$ (defined as $ f_\lambda ({\rm erg ~cm^{-2} s^{-1} nm^{-1}}) \propto \lambda^\beta$  for $\lambda > 120 $ nm) and  dust attenuation traced by $L_{\rm IR}/L_{\rm FUV}$  for local starburst galaxies observed by {\it IUE} and {\it IRAS}.  This  local starburst relation is now widely used to estimate dust attenuation in UV-selected galaxies. 
In the local universe, the $FUV-NUV$ colour  is used as a  proxy of $\beta$, even if the NUV band overlaps the bump area of the Milky Way dust extinction curve \citep[e.g.][]{conroy10}). \citet{johnson07} investigated the link between $L_{\rm IR}/L_{\rm FUV}$ (and therefore $A_{\rm FUV}$) and various colours for galaxies detected by {\it GALEX}, SDSS, and {\it Spitzer}. Their sample  represents both the red and blue sequence of the local universe and they found a  dependence of  $ L_{\rm IR}/L_{\rm FUV}$ on  D4000, i.e. on the age of the stellar populations. As described in section 3, our sample is strongly biased towards active star-forming galaxies, so we expect to find a smaller dispersion in the $A_{\rm FUV}$-colour relation than that found by  \citet{johnson07}.
  
We compare the dust attenuation parameter $A_{\rm FUV}$ obtained with our SED fitting method and two colours: the UV-only colour $FUV-NUV$ and the longer base-line colour $FUV-r$.\\
We first consider the $FUV-NUV$ colour and restrict our  analysis to sources with an error in $FUV-NUV$ that is smaller  than 0.2 mag . The values are reported in Fig~\ref{AFUV-colour} and  exhibit a  fairly weak  correlation (correlation coefficient R=0.59), a linear regression indicating that:
$$A_{\rm FUV} = 2.39~(\pm 0.20) \times (FUV-NUV) + 0.89~(\pm 0.13) ~~~(\sigma=0.83).$$
Several authors proposed a relation between $A_{\rm FUV}$ and $FUV-NUV$. \citet{Meurer99} obtained a linear relation between the slope of the UV continuum $\beta$ and $A_{FUV}$  for starburst galaxies, \citet{seibert05} found that the starburst relation overestimates $A_{\rm FUV}$ by $\sim 0.5$ mag  for a sample of galaxies not necessarily starbursting. These relations are   reported in Fig~\ref{AFUV-colour},  wheren when necessary $\beta$ is computed from the $FUV-NUV$ colour ($\beta = 2.23 (FUV-NUV)-2$ \citep{kong04}).  We find that the relation proposed by \citet{seibert05}  fits  our results  more closely than that of  \citet{Meurer99}  but,  in most cases, it  overestimates  the dust attenuation for a given $FUV-NUV$ colour.  \citet{salim07} derived a shallower relation by fitting UV-optical data only (without IR data) for normal blue galaxies.  Their relation is also reported in Fig~\ref{AFUV-colour} and appears consistent with our own regression.  In all cases the dispersion in  the data leads to a standard deviation around the regression line  of 0.8 mag.   The steepness of the relation between $FUV-NUV$ and $A_{\rm FUV}$ make the derivation of a dust attenuation from the $FUV-NUV$ color quite uncertain.  This weakness of the $FUV-NUV$ colour as a dust attenuation estimator  is also underlined by \citet{conroy10}. Using our regression formula  leads to an uncertainty of 0.5 mag on  $A_{\rm FUV}$ for an error of only 0.2 mag on the $FUV-NUV$ colour. Luminous infrared galaxies (LIRGs) with $L_{\rm IR} > 10^{11}~L_\odot$ (Fig~\ref{AFUV-colour}) are found to be closer to the starburst law than the bulk of the sample as already found by \citet{takeuchi10} and \citet{howell10}, but the dispersion in  their distribution remains large.

The long baseline $ FUV-r$ is now considered, where we again restrict the analysis to sources with an error smaller than 0.2 mag in this colour. A   weak  correlation (R=0.68) is found between $A_{\rm FUV}$ and $FUV-r$.  It is clear from Fig~\ref{AFUV-colour2} that the larger range of values obtained for $FUV-r$  compared to that for $FUV-NUV$ leads to a more reliable  estimate of dust attenuation for a similar uncertainty in the colours. 
A linear regression gives
$$A_{\rm FUV} = 0.65~(\pm 0.04)\times (FUV-r) +0.04~(\pm 0.16) ~~~(\sigma=0.76).$$
The RMS dispersion $\sigma$ is large reflecting the dispersion in the correlation. Nevertheless, the relation between  $A_{\rm FUV}$ and $FUV-r$ is much flatter than that between  $A_{\rm FUV}$ and $FUV-NUV$ implying an error of only 0.1 mag in   $A_{\rm FUV}$ for an error of 0.2 mag in  $FUV-r$.
\citet{johnson07} proposed several relations between $A_{\rm FUV}$, $NUV-r$, and  D4000. For consistency,   we prefer to compare  $FUV-r$ rather than $NUV-r$ with  $A_{\rm FUV}$. Nevertheless,  we have checked that  their relation  corresponding to D4000 = 1.25 (close to the mean D4000 found for our sample) is  compatible with our values of  $A_{\rm FUV}$ and $NUV-r$. \\
 
 Both colours   give only a crude estimate of $A_{\rm FUV}$ with an RMS dispersion of $\sim 0.8$  mag.  Since we have considered   galaxies actively  forming stars  the  $FUV-r$ colour is found to be a more reliable proxy than  $FUV-NUV$ of  dust attenuation in UV: only one UV band is necessary  and the relation with $A_{\rm FUV}$ is less sensitive to the photometric uncertainty than when  $FUV-NUV$ is considered. In the same way, \citet{johnson07} proposed to  use  the $NUV-3.6\mu$m colour.

\section{SFR calibrations}

\begin{figure}
 \includegraphics[width=8cm]{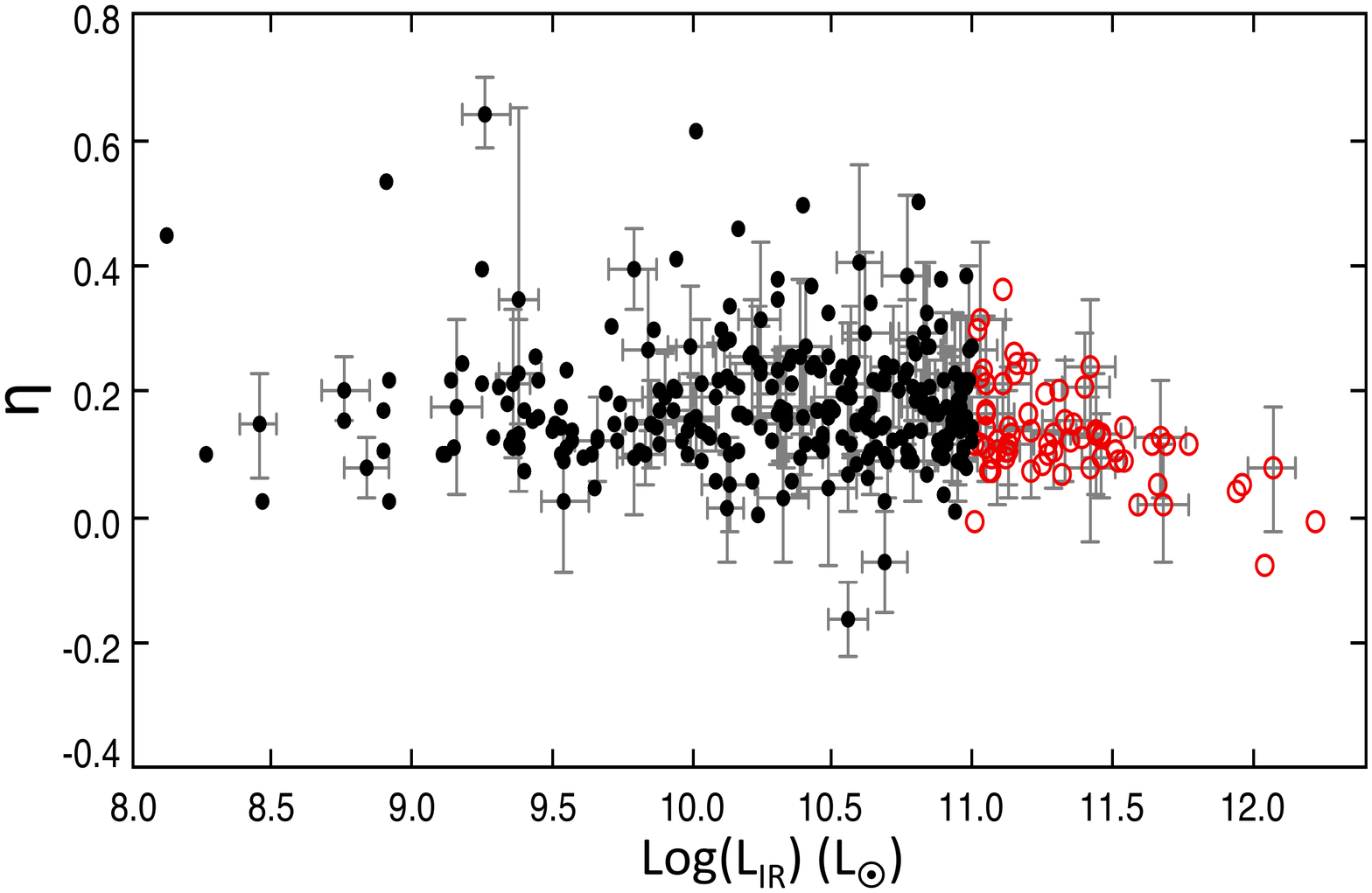}
 \caption{ Fraction of dust heating by old stars, $\eta$, plotted as a function of  $ L_{\rm IR}$, galaxies with $L_{\rm IR} > 10^{11}~L_\odot$ are plotted as empty red circles. Standard errors in  both output parameters given by CIGALE are plotted as error bars for 1/4 of the sample.}
      \label{eta-SFR}
\end{figure}

\begin{figure}
 \includegraphics[width=8cm]{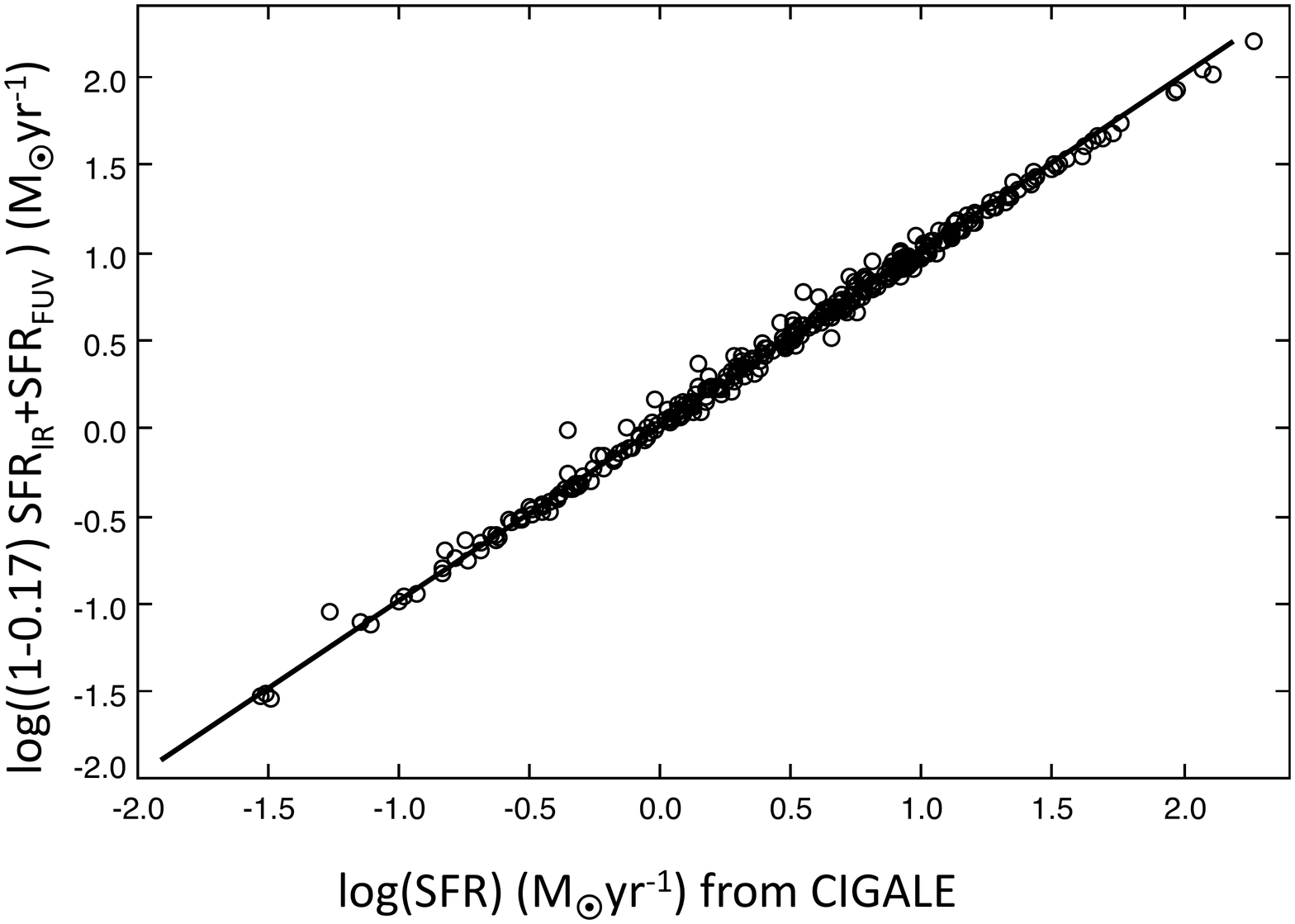}
  \caption{Star formation rates obtained from CIGALE (x-axis) versus the total star formation rates $\rm SFR_{tot}$ calculated with $\eta=0.17$ (y-axis).}
      \label{SFR}
\end{figure}

\begin{figure}
 \includegraphics[width=8cm]{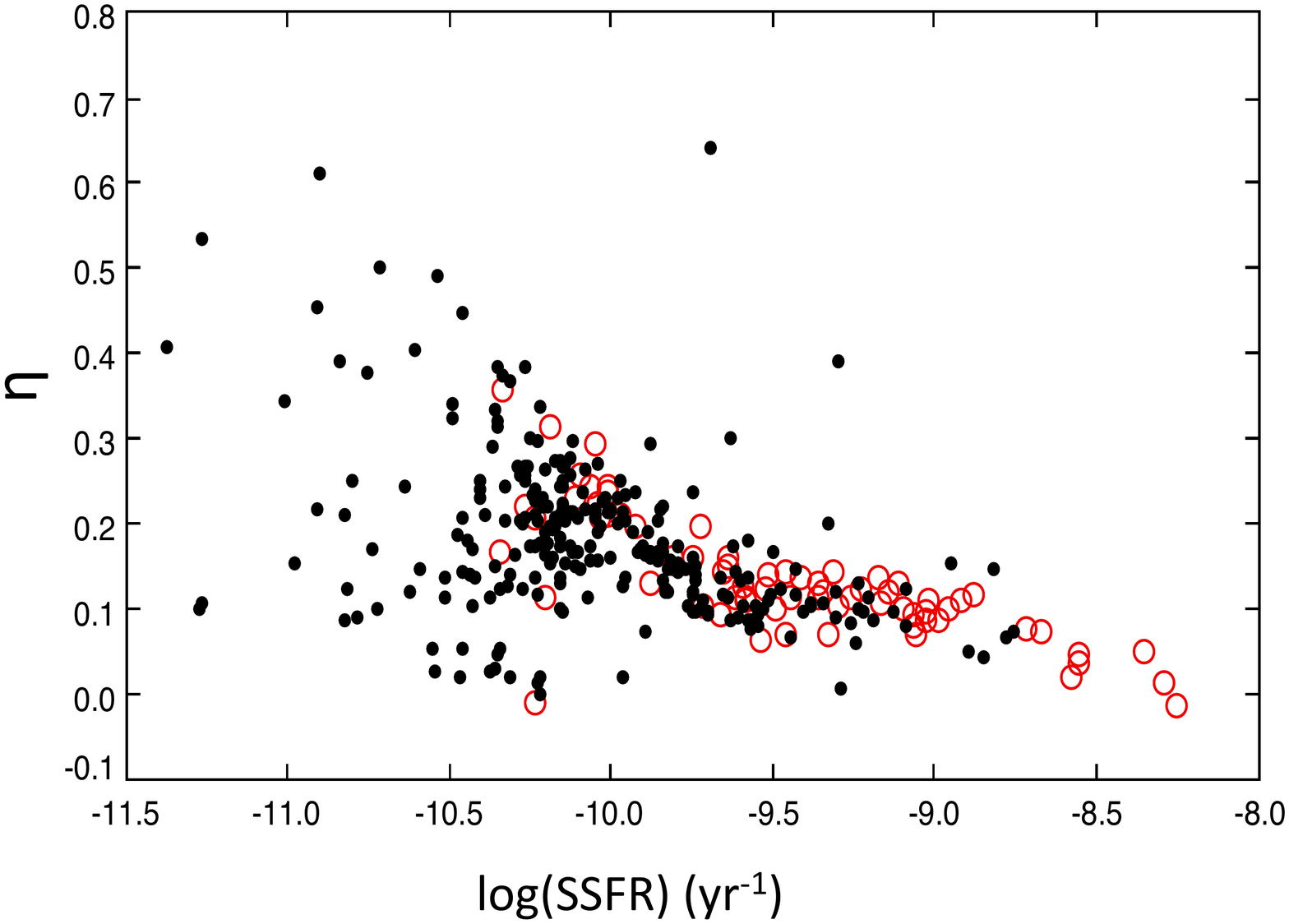}
 \caption{ $\eta$ plotted  as a function of  the specific star formation rate (SSFR), where galaxies with $L_{\rm IR} > 10^{11}~L_\odot$ are plotted as empty red circles.}
      \label{eta-SSFR}
\end {figure}

When both UV and IR fluxes are measured, a very efficient way to estimate SFR is to add the contributions of both types of emission to the total SFR \citep[e.g.][]{iglesias06,elbaz07}.
One must also account for dust heating by old stars, which is  not directly related to the current star formation and given by 
$$ \rm SFR_{tot}=(1-\eta) {\rm SFR}_{\rm IR}+{\rm SFR}_{\rm FUV},$$
 where $\eta$ represents the fraction of IR emission due to dust heating by old stars, $\rm SFR_{\rm IR}$ being  calculated  by converting  the  total IR luminosity $L_{\rm IR}$  into SFR assuming that all the light from stars is absorbed by dust and  $\rm SFR_{\rm FUV}$ is calculated from the observed UV luminosity $L_{\rm FUV}$. The relation between the IR and FUV luminosities and the corresponding SFRs are derived from population synthesis models assuming a particular  initial mass function and  star formation history. In this work, we use the ${\rm SFR}$ calibrations  from \citet{buat08} for  a constant ${\rm SFR}$ over $10^8$ years (the typical duration of the UV emission to reach a steady state) and a Kroupa initial mass function \citep{kroupa01}
: $\log({\rm SFR}_{\rm IR})_{\rm M_{\odot} yr^{-1}} = \log(L_{\rm IR})_{\rm L_{\odot}}-9.97$ and $\log({\rm SFR}_{\rm FUV})_{\rm M_{\odot} yr^{-1}} = \log(L_{\rm FUV})_{\rm L_{\odot}}-9.69$.  We prefer to combine SFRs rather than luminosities  to  avoid bolometric corrections in UV, which depend on the wavelength range assumed for the UV \citep{bell05}  the FUV being   taken to be monochromatic (at 153 nm) and directly calibrated in terms of the SFR. \\
In Fig~\ref{eta-SFR}, we  report the variation in $\eta$ as a function of $L_{\rm IR}$ for our sample, 
$<\eta> = 0.17 \pm 0.10$ for the whole sample. A small trend is found as a function of $L_{\rm IR}$ with  $<\eta> = 0.14 \pm 0.08$ for $L_{\rm IR} > 10^{11}~L_\odot$ (26 galaxies).  
Adopting $\eta=0.17$ leads to a very good agreement between  $\rm SFR_{tot}$ and the SFR estimated by CIGALE (Fig.~\ref{SFR}). This agreement implies that the timescales over which both SFRs are calculated are similar. As explained above, the calculation of $\rm SFR_{tot}$ assumes a constant SFR over $10^8$ years. The SFR deduced from the code is mainly governed by the second stellar population, which is produced at  a constant rate  over $t_2$ years (cf. Fig.~\ref{SFH_2}). Even if  $t_2$ is poorly  constrained (cf. section 3.2), its median value for the sample is 250 Myr, confirming that the SFR measured by CIGALE is consistent with a constant SFR over $\sim 10^8$ years.\\
The value of $\eta$ we find here is  lower than that found by \citet{hirashita03} ($\eta=0.4$). \citet{bell03} estimated the mean contribution of old stellar populations to $L_{\rm IR}$ to be $32\% \pm 16\%$ for $L_{\rm IR} < 10^{11}~L_\odot$ and $9\% \pm 5\%$ for $L_{\rm IR} > 10^{11}~L_\odot$, these values and those of the present work are compatible within their uncertainties.\\
\citet{dacunha10} estimated the fraction $f_\mu$  of total IR luminosity emitted by  dust in the diffuse ISM, stars being assumed to migrate in this ambient ISM $10^7$ years after their birth. This parameter is not directly comparable to $\eta$  for several reasons:  dust in the ISM is also heated by the young stars and a fraction of its emission is related to the star formation. We note that  we define $\eta$ by assuming a constant star formation rate over $10^8$ years and  stars in the ISM  younger than $10^8$ years  are also contributors to the current SFR. We therefore expect $f_\mu$ to be larger than $\eta$. Indeed, \citet{dacunha10} found median values of $f_\mu$ between 0.5 and 0.7 with a  dependence on the stellar mass and  the specific star formation rate. We  found  a very weak  positive correlation between  $\eta$ and $M_{star}$ and $\eta$ strongly correlates  with the SSFR (Fig.~\ref{eta-SSFR}) confirming the results of \citet{dacunha10}. This result is unsurprising since in galaxies that are very active in star formation the dust is mainly heated by newly formed stars.

\section{Conclusions}

We  have analysed the broad-band SEDs of 363 nearby galaxies from the UV to the far-IR by combining AKARI, SDSS, and GALEX surveys. The sample is close to representing one based on a far-IR selection. We fitted the SEDs with the code CIGALE that  performs a Bayesian analysis to deduce parameters related to the  dust attenuation and star formation  of each galaxy. An old (exponentially decreasing SFR) and a young (constant SFR)  stellar population are combined and the attenuated  stellar emission is related to the dust emission on the basis of energy conservation.  We have shown  that the detailed  star formation history is ill-constrained even when considering 11 photometric bands,   whereas global properties such as  the stellar mass, the total dust emission,  the star formation rate, the 4000 $\AA$ break, or the amount of dust attenuation are reliably estimated. \\
Our galaxies appear to be both massive and active in star formation. Different dust attenuations are  found for the young and  old stellar populations leading to a global dust attenuation in the V band $A_{\rm V}$ that is lower than that assumed for the young stellar population only, by a factor of 0.4  for galaxies with a 4000 $\AA$ break, D4000 $>$ 1.3,  and a factor of 0.7 for galaxies with D4000 $<1.3$  (D4000 being estimated from the SED fitting process). We have been unable to  constrain the presence  of a bump at 220 nm but have found  that a dust attenuation law that is slightly steeper than that proposed by \citet{calzetti00} more closely  reproduces our data. \\
The dust attenuation in the  FUV band of $GALEX$ is very well constrained by the $L_{\rm IR}/L_{\rm FUV}$ ratio. A polynomial calibration is provided and compared to other already published relations. 
We discussed the reliability of the $FUV-NUV$ and $FUV-r$ colours in estimating dust attenuation at FUV in the absence of IR data and find that both give only a very crude estimate of $A_{\rm FUV}$. The $FUV-r$ colour is more reliable  because the linear regression $A_{\rm FUV}$-$FUV-r$ is not as steep as that between $A_{\rm FUV}$ and $FUV-NUV$. \\
The SFR given by CIGALE is compared to the combination of SFRs estimated by adding the IR and the UV observed contributions. Both estimates agree remarkably well when a contribution of dust heating by old stars of 17$\%$ is assumed.
   
\begin{acknowledgements}
This work is based on observations with AKARI, a JAXA project with the participation of ESA.
VB, EG and DB have been supported by the Centre National des Etudes Spatiales (CNES).
TTT has been supported by Program for Improvement of Research 
Environment for Young Researchers from Special Coordination Funds for 
Promoting Science and Technology, and the Grant-in-Aid for the Scientific 
Research Fund (20740105) commissioned by the Ministry of Education, Culture, 
Sports, Science and Technology (MEXT) of Japan.
TTT  is partially supported from the Grand-in-Aid for the Global 
COE Program ``Quest for Fundamental Principles in the Universe: from 
Particles to the Solar System and the Cosmos'' from the MEXT.
\end{acknowledgements}


\begin{thebibliography}{}
\bibitem[\protect\citeauthoryear{Abazajian et 
al.}{2009}]{abaza} Abazajian K.~N., Adelman-McCarthy, J.K., Agueros, M.A. et al., 2009, ApJS, 182, 
543 
\bibitem[Balogh et al.(1999)]{balogh99}  Balogh, M. L.; Morris, S. L.; Yee, H. K. C.; Carlberg, R. G.; Ellingson, E. 1999, ApJ, 527, 54
\bibitem[Bell(2003)]{bell03} Bell, E. 2003, ApJ 586, 794
\bibitem[Bell et al.(2005)]{bell05} Bell, E.F.; Papovich, C., Wolf, C. et al. 2005, ApJ 625, 23
\bibitem[\protect\citeauthoryear{Boissier et 
al.}{2007}]{boissier07} Boissier S.,Gil de Paz, A., Boselli, A. et al. 2007, ApJS, 173, 524 
\bibitem[Boquien et al.(2009)]{boquien09} Boquien, M., Calzetti, D. , Kennicutt, R. et al., 2009, ApJ, 706, 553
\bibitem[Brinchmann et al.(2004)]{brinchmann04} Brinchmann, J., Charlot, S.; White, S. D. M.2004, MNRAS, 351, 1151
\bibitem[Buat et al.(2005)]{buat05} Buat, V., Iglesias-P\'aramo, J.; Seibert, M. et al. 2005, ApJ, 619, L51
\bibitem[Buat et al.(2008)]{buat08} Buat, V., Boissier, S., Burgarella, D. et al. 2008, A\&A, 483, 107
\bibitem[Burgarella et al.(2005)]{burgarella05} Burgarella, D.; Buat, V.; Iglesias-P\'aramo, J. 2005, MNRAS, 360, 1413
\bibitem[Calzetti et al.(1994)]{calzetti94}  Calzetti, D., Kinney, A.L., Storchi-Bergmann, T.1994, ApJ 429, 582
\bibitem[Calzetti et al.(2000)]{calzetti00} Calzetti, D., Armus, L., Bohlin, R. C. et al. 2000, ApJ, 533, 682
\bibitem[Cardelli et al.(1989)]{cardelli89} Cardelli, J.A., Clayton, G.C., Mathis, J.S. 1989, ApJ, 345, 245
\bibitem[Charlot \& Fall(2000)]{charlot00} Charlot, S., Fall, M. 2000, ApJ 539, 718
\bibitem[Chary \& Elbaz(2001)]{chary01} Chary, R., Elbaz, D. 2001, ApJ, 556, 652
\bibitem[Conroy et al.(2010)]{conroy10} Conroy, C., Schiminovich, D., Blanton, M.R. 2010, ApJ, 718, 184
\bibitem[Cortese et al.(2008)]{cortese08} Cortese C., Boselli, A., Franzetti, P. 2008, MNRAS, 386, 1157
\bibitem[da Cunha et al.(2008)]{dacunha08} da Cunha, E., Charlot, S., Elbaz, D. 2008, MNRAS, 388, 1595
\bibitem[da Cunha et al.(2010)]{dacunha10} da Cunha, E., Eminian, C., Charlot, S., Blaizot, J. 2010, MNRAS 403, 1894
\bibitem[Daddi et al.(2007)]{daddi07} Daddi E., Alexander, D.L., Dickinson, M. et al. 2007, ApJ, 670, 156
\bibitem[Dale et al.(2001)]{dale01} Dale, D., A., Helou, G., Contursi, A., Silbermann, N.A., Kolhatkar, S. 2001, ApJ, 549, 215 
\bibitem[Dale \& Helou(2002)]{dale02} Dale, D.~A. , Helou, G. 2002,  ApJ, 576, 159
\bibitem[Dale et al.(2007)]{dale07} Dale, D.~A., Gil de Paz, A., Gordon, K.D. et al.  2007,  ApJ, 655, 863
\bibitem[Elbaz et al.(2007)]{elbaz07} Elbaz, D., Daddi, E. , Le Borgne, D. et al. 2007, A\&A, 468, 33
\bibitem[Giovannoli et al.(2010)]{giovannoli10} Giovannoli, E., Buat, V., Noll, S., Burgarella, D., Magnelli, B. 2011, A\&A, 525, 150
\bibitem[Gordon et al.(2000)]{gordon00} Gordon, K. D., Clayton, G.C., Witt, A. N., Misselt, K.A. 2000, ApJ 533, 236
\bibitem[Iglesias-P\'aramo et al.(2006)]{iglesias06} Iglesias-P\'aramo, J., Buat, V., Takeuchi, T.~T., et al. 2006, ApJS, 164, 38
\bibitem[Iglesias-P\'aramo et al.(2007)]{iglesias07} Iglesias-P\'aramo, J., Buat, V., Hernandez-Fernandez, J., et al. 2007, ApJ, 670, 279
\bibitem[Hirashita et 
al.(2003)]{hirashita03} Hirashita, H., Buat, V., \& Inoue, A.~K.\ 2003, \aap, 410, 83
\bibitem[Inoue et al. (2006)]{inoue06} Inoue, A.K., Buat, V., Burgarella, D., Panuzzo, P., Takeuchi, T.,T., Iglesias-P\'aramo, J. 2006, A\& A, 370, 380
\bibitem[Johnson et al.(2007)]{johnson07} Johnson, B. D., Schiminovich, D., Seibert, M. et al.  2007, ApJS, 173, 377
\bibitem[Kauffmann et al.(2003a)]{kauffmann03a} Kauffmann, G., Heckman, T.M., White, S. et al. 2003, MNRAS, 341, 33
\bibitem[Kauffmann et al.(2003b)]{kauffmann03b} Kauffmann, G., Heckman, T.M., White, S. et al. 2003, MNRAS, 341, 54
\bibitem[Kawada et al.(2007)]{kawada07} Kawada, M., Baba, H., Barthel, P.D.  et al., 2007, PASJ, 59, 389
\bibitem[Kennicutt(2003)]{kennicutt03} Kennicutt, R.C., Armus, L., Bendo,  G. et al. 2003, PASP, 115, 928
\bibitem[Kong et al.(2004)]{kong04} 
Kong X., Charlot S., Brinchmann J., Fall S.~M., 2004, MNRAS, 349, 769 
\bibitem[\protect\citeauthoryear{Kroupa}{2001}]{kroupa01} Kroupa 
P., 2001, MNRAS, 322, 231 
\bibitem[Howell et al.(2010)]{howell10} 
Howell J.~H.,Armus, L., Mazzarella, J.M.  et al. 2010, ApJ, 715, 572 
\bibitem[Maraston(2005)]{maraston05} Maraston, C. 2005, MNRAS 362, 799
\bibitem[\protect\citeauthoryear{Martin et al.}{2005}]{martin05} 
Martin D.~C., Fanson, J., Schiminovich, D. et al. 2005, ApJ, 619, L1 
\bibitem[Meurer et al.(1999)]{Meurer99} Meurer, G.~R., Heckman, 
T.~M., \& Calzetti, D.\ 1999, \apj, 521, 64 
\bibitem[Noll et al.(2009a)]{noll09a} Noll, S., Pierini, D., Cimatti, A. et al. 2009a, A\&A 499, 69
\bibitem[Noll et 
al.(2009b)]{noll09b} Noll S., Burgarella D., Giovannoli E. et al.  2009b, A\&A, 507, 1793
\bibitem[Panuzzo et al.(2007)]{panuzzo07} Panuzzo P., Granato G.~L., Buat V., Inoue 
A.~K., Silva L., Iglesias-P\'aramo J., Bressan A., 2007, MNRAS, 375, 640 
\bibitem[Popescu et al.(2000)]{popescu00} Popescu, C.C., Misiriotis, A., Kylafis, N.D., Tuffs, R.J., Fischera, J. 2000, A\&A, 362, 138
\bibitem[Popescu et al.(2010)]{popescu10} Popescu, C.C., Tuffs, R.J, Dopita, M.A., Fischera, J.,, Kylafis, N.D.,  Madore, B.F.  2011, A\&A, 527, 109
\bibitem[Salim et al.(2007)]{salim07} Salim, S. Rich, M., Charlot, S. et al. 2007, ApJS, 173, 267
\bibitem[Saunders et al.(2000)]{saunders00} Saunders, W., Sutherland, W.J., Maddox, S.J. et al. 2000, MNRAS 317, 55
\bibitem[Seibert et al.(2005)]{seibert05} Seibert, M., Martin, D. C., Heckman, T.M. et al. 2005, ApJ 619, L55
\bibitem[Siebenmorgen \& Kr\"ugel(2007)]{siebenmorgen07} Siebenmorgen, R., Kr\"ugel, E. 2007, A\&A 461, 445
\bibitem[Silva et al.(1998)]{silva98} Silva, L, Granato, G.L., Bressan, A., Danese, L., 1998, ApJ 509, 103 
\bibitem[Schlegel et al.(1998)]{schlegel98}  Schlegel, D.,J., Finkbeiner, D.P., Davis, M. 1998, ApJ, 500, 525
\bibitem[\protect\citeauthoryear{Takeuchi et 
al.}{2010}]{takeuchi10} Takeuchi T.~T., Buat V., Heinis S. et al. 2010, A\&A, 514, A4
\bibitem[Tuffs et al.(2004)]{tuffs04} Tuffs, R. Popescu, C., Volk, H.J., Kylafis, N.D., Dopita, M.A. 2004, A\&A 419, 821
\bibitem[V\'eron-Cetty \& V\'eron(2006)]{veron06} V\'eron-Cetty , M.P., V\'eron, P. 2006, A\&A 455, 773
\bibitem[Walcher et al.(2008)]{walcher08} Walcher, J., Lamareille, F., Vergani, D. et al. 2008, A\&A 491, 713 
\bibitem[Walcher et al.(2010)]{walcher10} Walcher, J., Groves, B., Budavari, T., Dale,D. 2011, Ap\&SS 331, 1
\bibitem[Yamamura et al.(2010)]{yamamura} Yamamura, I., Makiuti, S., Ikeda et al.  2010, {\it AKARI}-FIS Bright Source Catalogue Release note Version 1.0
 \end{thebibliography}
\end{document}